\documentclass{emulateapj}
\usepackage{graphicx}

\newcommand{\msun}{${\rm M}_{\odot}$}

\newcommand{\lya}{Ly$\alpha$~}

\def\see{\mbox{$^{\prime\prime}$}}

\shorttitle{Observations of LyC emission from z$\sim$3--4 star-forming galaxies}
\shortauthors{Vanzella et al.}

\begin{document}

\title{On The Detection Of Ionizing Radiation Arising From Star--Forming Galaxies At
  Redshift $z\sim 3$--4 : Looking For Analogs Of ``Stellar Reionizers''}

%% from AASTeX v4.0. You can use \email to mark an email %address

\author{Eros Vanzella\altaffilmark{1},
Yicheng Guo\altaffilmark{2},
Mauro Giavalisco\altaffilmark{2},
Andrea Grazian\altaffilmark{3},
Marco Castellano\altaffilmark{3},
Stefano Cristiani\altaffilmark{1},
Mark Dickinson\altaffilmark{4},
Adriano Fontana\altaffilmark{3},
Mario Nonino\altaffilmark{1},
Emanuele Giallongo\altaffilmark{3},
Laura Pentericci\altaffilmark{3},
Audrey Galametz\altaffilmark{3},
S. M. Faber\altaffilmark{5},
Henry C. Ferguson\altaffilmark{6},
Norman A. Grogin\altaffilmark{6},
Anton M. Koekemoer\altaffilmark{6},
Jeffrey Newman\altaffilmark{7},
Brian D. Siana\altaffilmark{8}
}

\affil{$^{1}$INAF Osservatorio Astronomico di Trieste, Via G.B.Tiepolo 11,
  34131 Trieste, Italy} 
\affil{$^{2}$Department of Astronomy, University of
  Massachusetts, 710 North Pleasant Street, Amherst, MA 01003}
\affil{$^{3}$INAF Osservatorio Astronomico di Roma, Via Frascati 33,00040
  Monteporzio (RM), Italy}
\affil{$^{4}$National Optical Astronomy Observatory, PO Box 26732, Tucson, AZ 85726, USA}
\affil{$^{5}$UCO/Lick Observatory, University of California, Santa Cruz, USA}
\affil{$^{6}$Space Telescope Science Institute, Baltimore, USA,}
\affil{$^{7}$University of Pittsburgh, Pittsburgh, PA 15260 USA}
\affil{$^{8}$California Institute of Technology, CA, USA}
\email{vanzella@oats.inaf.it}



\begin{abstract}
We use the spatially--resolved, multi--band photometry in the GOODS South
field acquired by the CANDELS project to constrain the nature of candidate
Lyman continuum (LyC) emitters at redshift $z \sim 3.7$ identified using
ultra--deep imaging below the Lyman limit (1-sigma limit of $\approx$30 AB
in a 2\see~diameter aperture). In 18 candidates, out of a sample of 19 with
flux detected at $>$3-sigma level, the light centroid of the candidate LyC emission
is offset from that of the LBG by up to 1.5\see. We fit the SED of the LyC candidates 
to spectral population synthesis models to measure photometric redshifts and the 
stellar population parameters. We also discuss the differences in the UV colors between 
the LBG and the LyC candidates, and how to estimate the escape fraction of ionizing 
radiation ($f_{esc}$) in cases, like in most of our galaxies, where the LyC emission is 
spatially offset from the host galaxy. In all but one case we conclude that the candidate LyC 
emission is most likely due to lower redshift interlopers. Based on these findings, 
we argue that the majority of similar measurements reported in the literature need further 
investigation before it can be firmly concluded that LyC emission is detected. Our only 
surviving LyC candidate is a LBG at $z=3.795$, which shows the bluest $(B-V)$ color among
LBGs at similar redshift, a stellar mass of $M \sim 2 \times 10^9$\msun, weak interstellar
absorption lines and a flat UV spectral slope with no \lya in emission. 
We estimate its $f_{esc}$ to be in the range 25\%-100\%, depending on the dust and
intergalactic attenuation.
\end{abstract}

\keywords{galaxies: distances and redshifts - galaxies: high-redshift -
  galaxies: formation}

\section{Introduction}

The origin of the ionizing radiation, whether from galaxies or AGN,
responsible for the re-ionization of the universe at high redshift,
$z>6$, and for keeping it ionized at later epochs, is still poorly
constrained. The contribution of quasars to the hydrogen ionizing background
increases as we look back in time from $z=0$ to $z \sim 2$ as the peak of the quasar luminosity
function is approached (e.g., Fanidakis et al. 2011). Beyond redshift 2 their
contribution significantly decreases and the hydrogen photoionization rate is
most probably maintained by stellar emission (e.g., Siana et al. 2008; 
Faucher-Gigu\`ere et al. 2009; Haardt \& Madau 2011; but see Fiore et al. 2011).  
The star-forming galaxies at $z > 3$ are therefore the leading candidates to provide the 
remaining ionizing photons.
The severe IGM absorption at $z>4.5$ prevents us from observing directly the Lyman 
continuum (LyC) emission (Meiksin 2006; Inoue \& Iwata 2008). This is particularly
true during the re--ionization of the Universe ($z>7$). It is therefore essential
to identify diagnostics of the LyC emitters at lower redshift, $z \simeq 3-4$ (at
$\lambda_{rest}>1216$\AA), and infer if galaxies with these characteristics are
more common during the epoch of re--ionization (one Gyr earlier, or $z>7$). 

At $z\sim3-4$ the direct detection of LyC emission from galaxies is still 
difficult, because ionizing radiation is severely attenuated by neutral gas 
and by dust in the interstellar and circumgalactic medium (ISM and CGM) of the 
source itself, as well as by the intervening IGM. As a result, the number 
of reported detections of star--forming galaxies at high redshift, typically 
selected as Lyman--break galaxies or with
equivalent criteria, with LyC emission is very small (Shapley et al. 2006;
Iwata et al. 2009; Nestor et al. 2011; S06, I09 and N11, hereafter; 
Vanzella et al. 2010b, V10b, hereafter; Boutsia et al. 2011). The issue
is further complicated by the relatively high probability of finding faint
low--redshift galaxies at very close angular separation, i.e., $\sim$ 1 arcsec or
less, from a high--redshift galaxy with brighter apparent magnitude 
(Vanzella et al. 2010a, V10a, hereafter).  
These interlopers, for which spectroscopic
identification is not viable, can be wrongly interpreted as spots of LyC emission
from the host galaxy if high--angular resolution multi--band photometry is not
available, as is often the case, to recognize their nature.

Recently, searches of LyC emitters with deep imaging below the Lyman limit
have yielded samples of candidates at redshift z$\sim 3$ where the region of
emission of the candidate LyC ionization radiation is spatially offset from
that of the non--ionizing rest far--UV light (dubbed ``nLyC'' in what
follows), namely the main body of the galaxy typically observed between 
1500\AA~and 2000\AA. The displacement of these putative LyC--emitting ``blobs'' is
generally less than one arcsec, but in some cases values as high as 2$''$ have
been reported (I09; N11), corresponding to separation in the range of
several kpc to a few tens of kpc. In other words, the candidate LyC emission
would come from regions of the galaxies that are well separated from the
center, which is where most of the LyC photons are produced, and in some cases
so far from it to be classified as separate sources. If confirmed this
would provide important empirical constraints on how LyC escapes from
galaxies. Future observations with the new HST/WFC3 in the LyC rest-frame at
$z\sim3$ will help exploring this issue further. In the meantime, however,
motivated by the potential importance if the above findings are confirmed, we
use spatially resolved, multi--band photometry from {\it HST} to constrain the
nature of a sample of LyC emitter candidates that we have selected in the
GOODS South, one of the fields targeted by the CANDELS project (Grogin
et~al. 2011; Koekemoer et~al. 2011) from ultra--deep U--band imaging below the
Lyman Limit of star--forming galaxies with known spectroscopic redshift 
$z\ge 3.4$. We also report on the only one Lyman Break Galaxy in our sample
at $z=3.795$ ($Ion1$, hereafter) with a clear LyC emission.

\section{Analysis of a Sample of Candidates LyC Emitters in GOODS--South}

We have selected a sample of LyC emitter candidates from ultra--deep imaging
in the U--band of the GOODS--South field (V10b). 
The U--band images, which are described in Nonino et~al. (2009), were aimed 
to probe the spectral region below the Lyman limit of these
galaxies for possible emission of ionizing radiation. The images are very
deep, reaching 
1-$\sigma$ flux upper limit of about 30 mag (AB) for an
unresolved source within a circular aperture of 2\see\ diameter.

A relatively
large number of sources, selected as Lyman--break galaxies,
have secure redshift identification in the GOODS-S field
(Vanzella et~al. 2008, 2009; Popesso et~al. 2009; Balestra et~al. 2010).
As described in V10b, we start from a spectroscopic sample 
of 122 B--band dropouts at $3.4\le z\le 4.5$ selected by Giavalisco et~al. (2004)
from the GOODS/ACS images for which we have robust redshifts,
i.e., Quality Flag QF=A.
The lower limit of the redshift range is the lowest value such that the
912\AA\ Lyman limit is outside, redward of the system throughput in the U
filter, while the upper is chosen because the rapid increase of opacity of the
IGM produces too small transparency at higher redshift to make analysis of
$z>4.5$ galaxies useful (see V10b). 7 out of 122 B--band dropouts are detected
in the Chandra $4~Msec$ images in the CDF-S (Xue et al. 2011) and classified
as AGN.
Among these 122 sources we have flagged as potential LyC emitter
candidates the 32 of them which have flux detection at the $2\, \sigma$ level
or larger within a circular aperture of 1.2\see\ diameter
in the U--band image. 
In this work we discuss in detail these U-band emitters (named $Uem$ hereafter).

As we noticed in V10b, in 28 of the 32 sources the U--band emission, i.e., the
candidate LyC light, is spatially offset relative to the centroid of the
rest--frame far--UV light of the LBG (the nLyC at wavelength around $\approx 1500$ \AA). 
A visual inspection of the {\it HST}/ACS images showed that in 9
of the 28 sources the U--band flux light can unambiguously be explained as
coming from the outer isophotes of nearby sources, which most likely are
foreground interlopers (e.g., see Figures 4 and 5 of V10b). We have eliminated
these galaxies from the sample of LyC candidates. The remaining 19 candidates
are listed in Tab.~1 and are analyzed here.
They have counterparts in
the ACS and WFC3 images which are close to the targeted host LBG 
with angular separations in the range
0.4\see--1.9\see. Image cutouts for these sources are shown in V10b, here we show in 
Figure~\ref{fig1} and the Appendix the more critical cases where the angular separation is 
smaller than 1\see.
All but one of the U--band
sources are fainter, in AB magnitudes, than their ACS optical (rest UV) nLyC
counterparts. Tab.~1 reports the ratio of $f1500/f_{LyC}$ for the 19 objects.

The remaining 4 U--band detections of the initial sample of 32 are
co--spatial with the ACS and WFC3 images of the host sources, in the sense that
the centroid of the U--band light falls within the errors ($\sim$ 0.1\see) 
on the location derived from the optical HST/ACS images. As discussed in
V10b (and reported in their Table~3), 3 out of 4 are detected in the 4 Msec X--ray Chandra images 
(0.5-8 keV), and also show typical signatures of
AGN in their optical spectra, such as high ionization emission lines (e.g.,
\ion{C}{4}, \ion{N}{5}). The forth source, which we call {\em Ion1}, is our only
robust candidate $stellar$ LyC emitter and we will describe it further in Sect. 6.

\subsection{The Effect Of The Intergalactic Attenuation}

The method we are adopting here is based on an intermediate--band filter 
($FWHM=350$\AA, U--band), and is similar to the typical narrow--band imaging only in the cases where
$3.4<z<3.5$ (9 out of 19 LBGs), i.e. the $\Delta \lambda$ between
the Lyman limit of the galaxy and the red cut-off of the filter is minimal.
In this case, the greater depth of our image compensate for the
larger noise due to the broader bandpass. For example, if compared with the
NB3640 narrow--band filter used by N11 ($FWHM=100$\AA), 
the depth of our U-band image nearly exactly compensate for the differet filter setup.

\begin{figure}
\epsscale{1.0}
\plotone{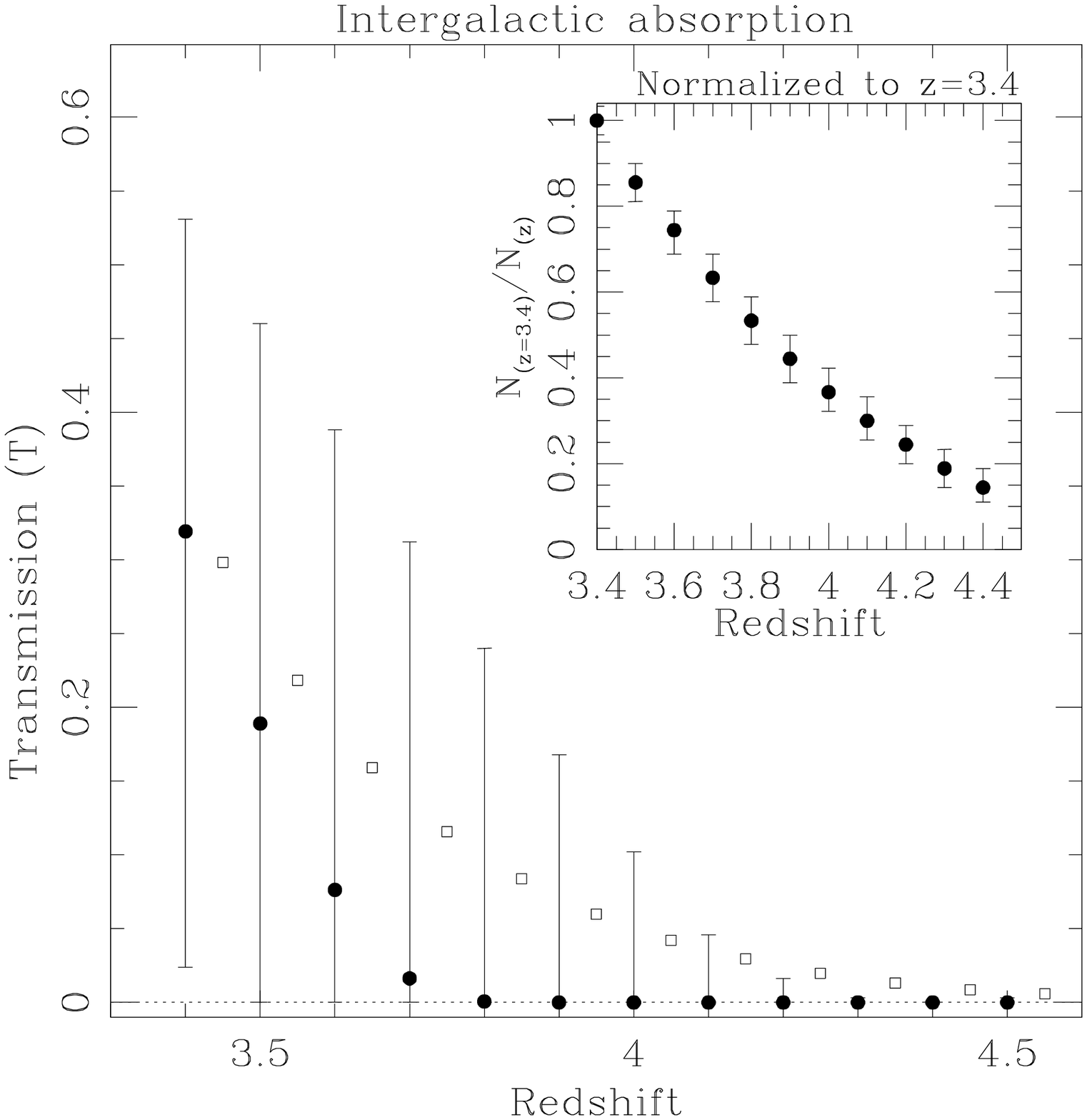} %NewTransmission.eps} %f1new.ps} 
\caption{{\bf Main box}: Transmission 
%averaged over the wavelength range of the LyC(880--910\AA) (open circles) and 
convolved with the VLT/VIMOS U-band filter 
(filled circles) as a function of source redshift. The filled circles and vertical error bars
indicate the median value and central 68\% range of the transmission for the
10,000 lines of sight generated with the new Inoue et al.~(2011) simulations. Open squares are
the averages calculated over the same lines of sight (shifted by dz = 0.05 to
the right for clarity). Clearly, the VLT/VIMOS U-band probes progressively shorter
wavelengths as redshift increases, with the effect of lowering the transmission.
{\bf Inner box}: The fraction of recovered sources by the U-band imaging (with $S/N>2$) as a function 
of the increasing IGM attenuation (redshift), normalized to the $z=3.4$ case (see text for details).
 \label{figT}}
\end{figure}

\begin{figure}
\epsscale{1.0}
\plotone{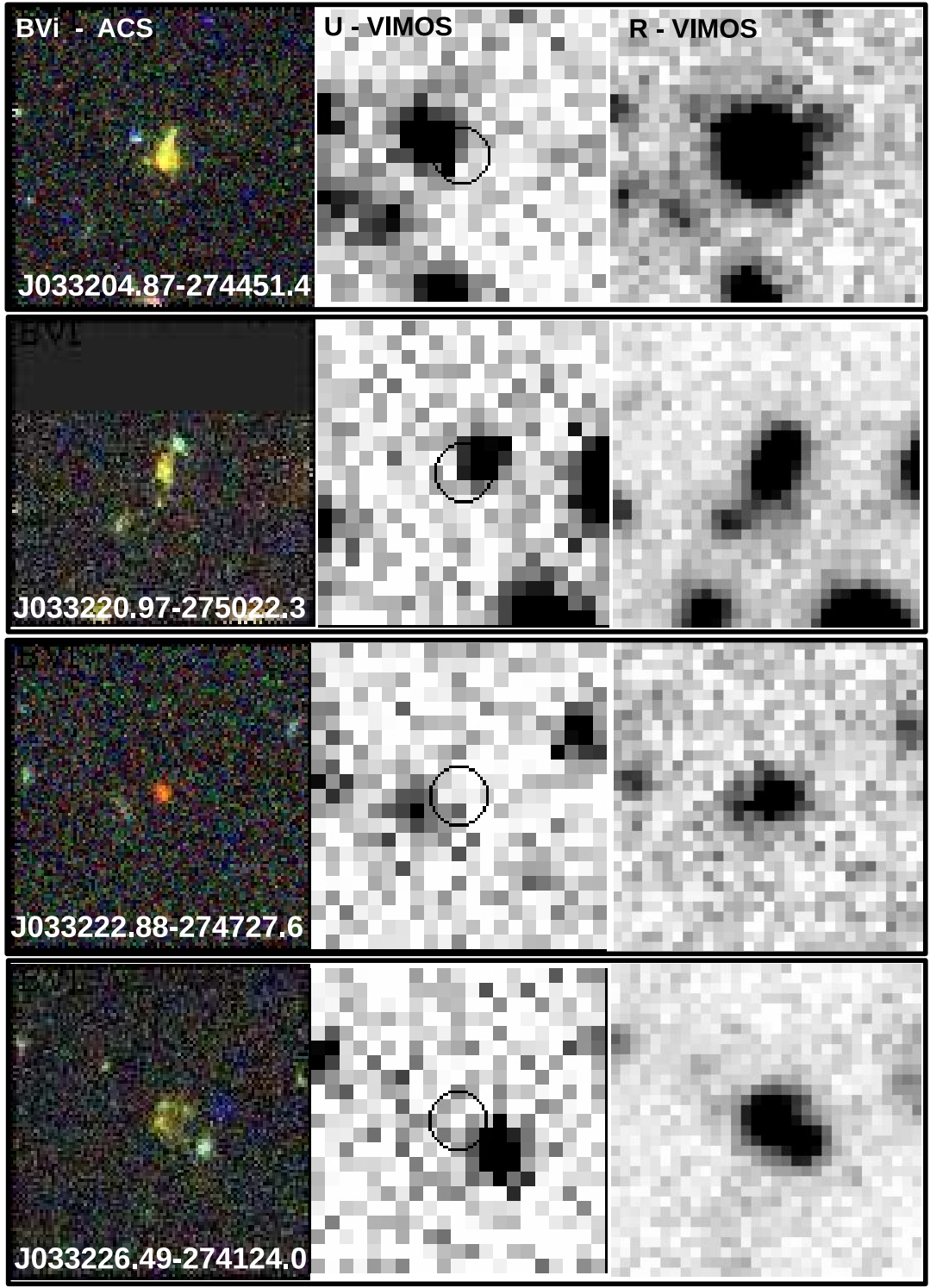} %{Figure_less1arcsec.eps} %f1new.ps} 
\caption{Examples of 4 LBG (out of 19) with the closest offset U--band detections
(and see Figure~\ref{fig7}). 
If the U--emission arise from a region that is at the same redshift as that of the LBG, 
than it would be genuine LyC emission.
The angular separation between the main targets (LBG marked in the U--band with 1.2\see diameter circles)
and the fainter counterparts that generate the U--band emission is smaller than
1.0\see~(see Table~\ref{tab1}). The color ACS/BVI, VIMOS/U and VIMOS/R images are shown
for each case, from left to right, respectively, and the GOODS ID is reported for the 
targeted LBG. The box sizes are 6.3\see~on a side.
 \label{fig1}}
\end{figure}

Differences between the methods arise if we consider, as in our approach,
a variable redshift. Indeed, the LyC region probed by our
filter is redshift dependent, e.g., it is $\lambda_{rest} < 909,800,741$\AA~at 
$z=3.4,4.0,4.4$, respectively. We have investigated the effect of the IGM
attenuation as a function of redshift by running MC simulations as in V10b
(where the IGM attenuation is extensively taken into account).
To this end, as a reference, we adopt a magnitude of $i_{775}=24$ and a 
ratio $f_{1500}/f_{LyC}=7$ (V10b, Siana et al. 2007).
We then apply 10000 different IGM transmissions convolved 
with the U-band filter and add photometric noise to the estimated flux in the U-band 
(see (V10b)). In these conditions, at $z=3.4$ we retrieve $82_{-4}^{+4}\%$ of the sources in their LyC 
(U-band) at $S/N>2$ ($88_{-4}^{+3}\%$ if $i_{775}=22$).
As redshift increases, the fraction of recovered sources 
decreases because the IGM attenuation (see Figure~\ref{figT}).
The inner box of Figure~\ref{figT} shows how the IGM affects the fraction of recovered sources 
as a function of redshift (normalized to the $z=3.4$ case).

It is worth noting that the majority of the sources analyzed here are at $z<3.8$ (15 out of 19). 
Moreover, two examples at relatively high redshift have recently been reported: a LBG at $z=3.8$ 
({\em Ion1}, also discussed in this work) and a AGN at $z=4.0$ with $i_{775}=26.09$,
for which a LyC emission at $\lambda_{rest}<830$\AA~and 
$\lambda_{rest}<800$\AA~is detected at S/N of 5.2 and 3.3 by our deep U-band imaging, respectively (V10b).
This suggest that despite the low average IGM transmission at $z\gtrsim 3.8$, its
stochastic behavior makes the variance to be relatively large. 
Indeed, at fixed redshift, the distribution of the IGM transmissions (U-band convolved) is 
asymmetric with an extended tail 
toward high values (Inoue \& Iwata 2008). The reason is that the Lyman continuum absorption
by the IGM is very stochastic because it is related to the presence
of relatively rare Lyman limit systems (LLS) or damped Lyman $\alpha$ system (DLA),
having $NHI>10^{17}~cm^{-2}$. 
With a Lyman limit system, the transmission is suddenly
cut down at the corresponding wavelength.
Conversely, without a LLS (or DLA system) 
near the source, we can expect a significant
transmission even far below the source Lyman limit (see Inoue \& Iwata 2008).

Having this in mind, we decided to keep the whole sample up to $z\sim 4.4$ 
in the analysis performed in the present work.
If we could establish that the U--band selected sources observed in proximity
of their LBG were at the same redshift of their companions then we would safely
conclude that they are LyC emitters. Before to investigate the nature of
their redshift, in the following section we briefly recall
the issue of the contamination by lower redshift sources randomly placed 
at small angular separations from the higher redshift LBG.

\section{The Occurrence Of Foreground Contamination}

The likelihood that a foreground interloper located in the vicinity 
of a LBG is responsible for the detection in the U band
increases with redshift of the LBG, with the angular separation and the
quality and depth of the observations. 
Siana et al. (2007), I09 and N11 
calculated analytically the probability that candidate LyC emission from a LBG 
is due to an interloper, given the characteristics of the data (quality and depth).
V10a performed the same calculations and ran Monte-Carlo simulations to 
quantify this effect.

The displacement of the U--band light relative to that at redder wavelengths,
as measured from the light centroid in the U--band and ACS images, for the
sample 19 galaxies discussed here, is in the range $0.4<\Delta\theta<1.9$
arcsec; in all cases there always is a counterpart to the U--band source in
the {\it HST}/ACS and WFC3 images. We calculated that the number of sources
observed in the two annular bins with radii 0\see-1\see~and 
1\see-2\see~from the LBG centroid in the ACS $z_{850}$--band
images is equal, within the errors, to the expectations for foreground
galaxies at increasing separations. That is, given the number counts
and assuming a uniform distribution, the fraction of intercepted 
foreground sources increases with the area of the annulus considered.

One of the parameters adopted in V10a was the seeing (the PSF of the images), 
strictly related to the possibility to deblend close sources and related to the
probability of superposition, which increases for the worst seeing conditions. The fact that 
the LyC emission could be intrinsically offset from the main galaxy (as reported in N11 and I09) 
further complicates the interpretation, if the redshift of these potential emitters is not known.
This is still true if high spatial resolution images are available and the sources 
are well separated, i.e., in absence of the redshift information an intrinsically
offset LyC emission is fully compatible with the emission of a lower redshift object. 

On the one hand, the effects of this foreground contamination can be corrected 
statistically. On the other hand, it is worth investigating carefully
each LyC candidate, since any consideration about the mechanisms
that allow the LyC photons to escape primarily depend on the 
reliability of the LyC detection.
Therefore we now turn to the discussion of observational evidence 
that will help us to constrain the redshift, and
hence the nature, of these sources in our sample (Sect. 4).

\section{Constraining the redshift of the candidates LyC emitters in the GOODS-S sample}

\subsection{The Escape Fractions}

In this section we calculate the absolute and relative Lyman continuum 
escape fractions (defined below) of our potential
LyC emitters assuming that they are at the same redshift of the LBG.
Since $f_{esc}$ and ($f_{esc,rel}$) have to obey to clear limits, this in turn
puts constraints on other quantities, in particular the redshift.
In order to do that, we have to first define the relation among various quantities 
and calculate the escape fraction where spatially offset LyC and nLyC 
emissions are present.

\subsubsection{Escape Fraction From a Morphologically Resolved Lyman Continuum}

Following Siana et al. (2007) the observed flux ratio between the
1500\AA~and the Lyman continuum is affected by several factors and is
expressed as :

\begin{eqnarray}\label{eq:radio_tot}
\left(\frac{f1500}{f_{LyC}}\right)_{OBS} &=& \left(\frac{L1500}{L_{LyC}}\right)_{INT} \times 10^{-0.4(A1500-A_{LyC})}
\times \nonumber \\
&\times&  e^{\tau_{HI,IGM}(LyC)} \times e^{\tau_{HI,ISM}(LyC)}, \label{eq:fratio}
\end{eqnarray} 

{\noindent}where LyC is the wavelength at which the Lyman continuum is observed,
$(L1500/L_{LyC})_{INT}$ is the intrinsic luminosity density ratio,
($A_{1500}-A_{LyC}$) is the differential dust attenuation (in magnitudes), 
$\tau_{HI, IGM}(LyC)$ is the line-of-sight opacity of the IGM for LyC photons 
(and transmission is $T_{LyC}^{IGM}=e^{-\tau^{\rm IGM}_{LyC}}$), and 
$\tau_{HI, ISM}(LyC)$ is the optical depth of the Lyman continuum absorption
from H\,{\sc i} within the observed galaxy's interstellar medium, ISM 
(whose transmission is defined as $T_{ISM}^{HI}=e^{-\tau_{HI, ISM}(LyC)}$).

The {\it relative} fraction of escaping LyC photons relative to
the fraction of escaping nLyC (1500\,\AA) photons is obtained rearranging the
above equation :
\begin{equation} f_{\rm esc,rel} = \frac{(L1500/L_{LyC})_{\rm int}}{(F1500/F_{LyC})_{\rm obs}} \exp(\tau^{\rm IGM}_{LyC}), \label{eq:f_esc1}
\end{equation}

{\noindent} it compares the observed flux density ratio
(corrected for the IGM opacity) with models of the ultraviolet spectral energy
distribution of star-forming galaxies.  If the dust attenuation $A_{1500}$ is
known, $f_\mathrm{esc,rel}$ can be converted to $f_\mathrm{esc}$ as
$f_\mathrm{esc} = 10^{-0.4A_{1500}} f_\mathrm{esc,rel}$ (e.g., Inoue et
al. 2005; Siana et al. 2007). Again, from the above equations, 
$f_{\rm esc}$ can be written as:

\begin{equation} f_{\rm esc} = \textrm{exp} \left[-\tau_{HI, ISM}(LyC)\right] \times 10^{-0.4(A_{LyC})}, \label{eq:fescT}
\end{equation}

{\noindent} the two factors on the right side have values in the range [0--1].
Clearly, their product, i.e., $f_\mathrm{esc}$, cannot be greater than 1.

It has been recently argued by N11 that due to the details of the radiative
transfer and sources of the corresponding photons, any \lya emission and
escaping LyC flux will not be necessarily co-spatial with either each other or
with the bulk of the rest-frame UV flux in a given galaxy.  This is true when
considering \lya emission, which typically arises from backscattering of moving
hydrogen gas that can be spatially decoupled from the ionizing sources.
The observed light centroids (barycenter) of LyC and nLyC can be displaced
in the case in which the LyC arises from a sub-region of a larger area,
however, the $local$ emission is not spatially shifted,
i.e., if the ionization radiation is measured in some sub-region, then the 
nLyC radiation is expected to be detected too (typically with brighter magnitude).
{\it The crucial point here is that the two quantities $(F1500)_{\rm  obs}$
and $(F_{LyC})_{\rm obs}$ must be measured in the same spatial
(i.e., physical) region, where the ionizing and non-ionizing radiation arise.}

\begin{figure}
\epsscale{1.0}
\plotone{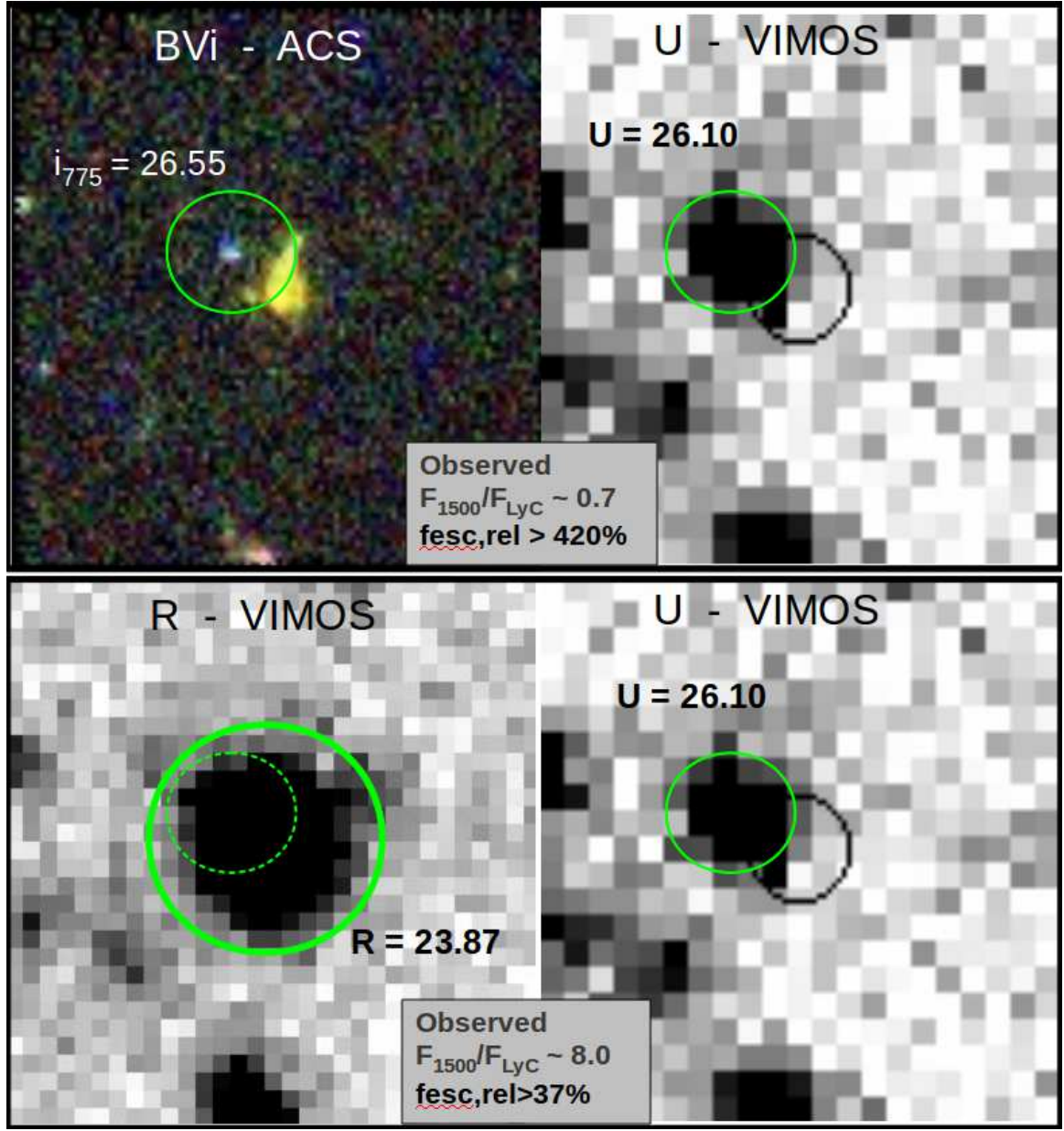} %Fig_exemple_region.eps} %f1new.ps} 
\caption{Illustrative example of the measurement of the observed flux
ratio between LyC e nLyC for the same system, but in different image 
spatial resolutions.
In the top panel the fluxes are measured and compared within the same 
regions (indicated with green solid circles), allowed by the high angular 
resolution available from HST. 
In the bottom panels the same is shown for ground-based observations,
in which the magnitude of the LBG (that includes the fainter offset blob) 
is wrongly adopted as $f1500$, as it has been done in recent works. 
The two ratios differ by more than a factor of 10 (see text and 
Table~\ref{tab1} for details) like their $f_\mathrm{esc,rel}$.
The solid green circles of the top panels are reported also
in the bottom-left (dashed circle) and right panels.
The black solid circles indicate the position of the LBG.
In this example we consider negligible the difference between
the nLyC fluxes probed by the $i_{775}$ and $R$ bands.
 \label{fig5}}
\end{figure}

Conversely, recent works have performed measures with the aim to include both
the fluxes $(F1500)_{\rm obs}$ and $(F_{LyC})_{\rm obs}$ by
enlarging the apertures (I09) or by deriving $total$ fluxes from the images regardless of the
misalignment (e.g., SExtractor MAG-AUTO in $U$ and $R$ bands, as in N11),
resulting in measures within different apertures size and/or shape.  
In this way, the measured observed ratio $(F1500/F_{LyC})_{\rm obs}$, and consequently the
$f_\mathrm{esc}$ quantity, are strongly biased. In particular
the resulting $f_\mathrm{esc}$ would be severely underestimated, because the
correct $(F1500)_{\rm obs}$ value would be smaller. Figure~\ref{fig5} shows an
illustrative example in the GOODS-S field (also discussed below and reported
in Tab.~1) in which the observed ratio has been calculated with and without taking 
into account of the offset emission, top and bottom panels, respectively.
Under the same assumptions (intrinsic luminosity ratio and IGM transmission), 
the former method (top panel) produces an $f_\mathrm{esc}$ that is
$\sim 10$~times lower than the latter (bottom panel).

The consequence is that if the estimated 
$f_\mathrm{esc}$ exceeds the value of 100\% then some other quantity has to be revised.
It is even more significant if this happens under conservative
assumptions of $(L1500/L_{LyC})_{\rm int}$, dust and IGM attenuation.

If, from one hand, the $f_\mathrm{esc}$ has to be less than 100\%, 
the constraints can be even more stringent if $f_\mathrm{esc,rel}$ has to be
less than 100\%. 
Indeed from Eq.~\ref{eq:fratio} and ~\ref{eq:f_esc1} it turns out that :

\begin{equation} 10^{-0.4(A_{1500}-A_{LyC})} \times f_{\rm esc,rel} = \textrm{exp} \left[-\tau_{HI, ISM}(LyC)\right] \label{eq:fesc2}
\end{equation}

{\noindent} In order to have the transmission of the 
ISM correctly in the range [0--1] (right side), the $f_{esc,rel}$ has to be less than $10^{0.4(A_{1500}-A_{LyC})}$,
which in turns is a quantity less than one (see Siana et al. 2007, their Figure~2).
Therefore also the relative escape fraction has to be less than one.

\subsubsection{Anomalous Escape Fractions In Our Sample}
Following the discussion of the previous section, 
we have estimated $f_\mathrm{esc,rel}$ for our sample in
the correct way, i.e., in same spatial regions where the $U$--band emission 
is observed, assuming that the U--band sources are at the same redshift
as the LBG, i.e., they are LyC emitters.  Following Eq.~\ref{eq:f_esc1} we calculate the
$f_\mathrm{esc,rel}$ by assuming an intrinsic ratio
$(L1500/L_{LyC})_{\rm int}=3$ (Shapley et al. 2006) and the maximum IGM transmission at the redshift
of the LBG.\footnote{$T_{LyC}^{IGM}$ has been calculated following the Inoue
  \& Iwata (2008) prescription convolved with the VIMOS/U band filter, as
  described in V10b; here we include in the calculations the recent statistics
  of the LLS provided by Prochaska et al. (2010) and Songaila \& Cowie (2010),
  that increase slightly $T_{LyC}^{IGM}$ (see Inoue et al. 2011).
  Transmissions at the redshift of the LBG have been calculated on 10000
  random line of sights.}
The intrinsic ratio is also justified by the fact that
the sources show an UV spectral slope (from their $(i-z)$ color) similar or 
redder than their LBGs.
Therefore the adopted ratio is a conservative assumption.

The $f_\mathrm{esc,rel}$ values are reported in the last column of Table~1.
All but one U--band LyC candidates have $f_\mathrm{esc,rel}$ larger than
100\%, that would implies a less dust attenuation for shorter wavelengths,
i.e., $A_{LyC} < A_{nLyC}$, in contrast to any dust extinction law.  This is
even more evident if we adopt the ratio $(L1500/L_{LyC})_{\rm int}=7$
(e.g., Siana et al. 2007; V10b), that increases the $f_\mathrm{esc,rel}$ values
of Table~1 of a factor $7/3$. The one case with $f_\mathrm{esc,rel}<100\%$ has
inconsistent photometric redshift and UV colors as described below.

\subsection{The UV Colors}

The $B-V$ color of galaxies at the redshift considered here largely depends
on the cosmic opacity of the IGM (see Madau 1995), specifically due to the 
effects of the \lya forest. Given the spatial correlation scale of the IGM,
if the U-band companions are at the same redshift as their associated LBGs,
their $B-V$ colors should therefore be similar.

\begin{figure}
\epsscale{1.0}
\plotone{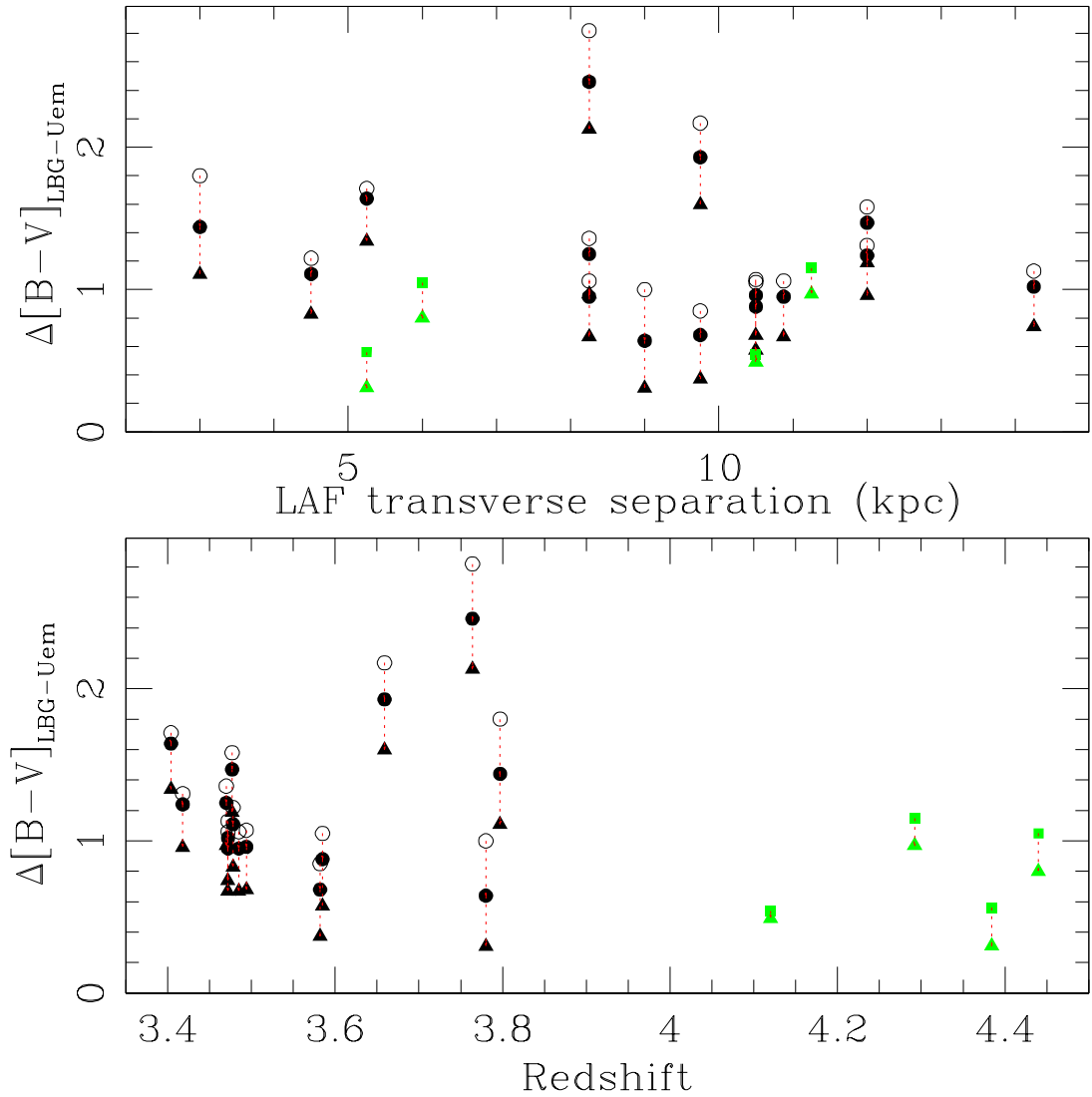} %fig4_transvIGM.eps}
\caption{Differences between the UV colors of LBGs and the U-band emitters 
($\Delta(B-V)=(B-V)_{LBG}-(B-V)_{Uem}$) as a function
of redshift (bottom) and physical transverse separation of the pairs (LBG-$Uem$) 
at the mean redshift of the Lyman-alpha forest (top). Open circles are the observed 
differences, whose typical error is $<0.2$ (see Tab.~\ref{tab1} and their S/N 
in the $B_{435}$--band).
Solid circles are the same quantity after correcting the colors by running MC simulations: 
a Lyman continuum emission with $f_{esc}=1$ has been inserted in the $B_{435}$-band of the $Uem$. 
Filled triangles are the colors derived from the same MC simulations by assuming a conservative 
IGM transmission of 100\% (see text for details).
Green symbols represents the $\Delta(V-i)$ for the pairs with LBG at $4.0<z<4.5$. In these cases the
$V_{606}$-band is more suitable than the $B_{435}$-band and includes only the IGM 
at $\lambda > 912$\AA~(no LyC is included in the $V_{606}$ filter (see text)).
\label{fig4}}
\end{figure}

Figure~\ref{fig4} shows the difference of the observed $(B-V)$ color $\Delta(B-V)$ 
between the LBG and their companions LyC candidates (empty circles,
$\Delta(B-V)=(B-V)_{LBG}-(B-V)_{Uem}$) as a function of the
transverse separation (physical) calculated at the mean redshift of the
\lya forest. The same is shown for the four $4.0<z<4.5$ galaxies
for which the $\Delta(V-i)$ has been adopted, more suitable than the $(B-V)$ color
in probing the IGM decrement. For all, the typical error in the color 
is $\lesssim$0.1 mag, because these sources are detected at more than 10-sigma 
in each band. As the figure illustrates, the LyC candidates are
much bluer than their companion LBGs, implying either a very high fraction of
escaping ionizing radiation and/or an high transmission of the IGM, or
more likely as we are about to argue, that their redshift is significantly 
lower than that of the LBG, in agreement with the previous result and 
the photometric redshift analysis (reported in the next section). 
This is seen through a Monte Carlo simulations in
which we have simulated $f_{esc}=1$ for the LyC candidates in the UV
rest--frame continuum probed by the $B_{435}$-band and we have conservatively assumed
that the IGM transmission at $\lambda < 912$ \AA\ is the same as that at
$912 < \lambda < 1215$ \AA\ (for the simulations we used the IGM
transmission by Inoue et~al. (2008, 2011) and the SB99 UV templates with age 100 Myr,
continuous star formation and metallicity z=0.004 and Salpeter IMF). We can
then correct the observed $(B-V)$ color of the LyC emitters and recalculate the
$\Delta(B-V)$, which we plot in the
Figure~\ref{fig4} as filled circles (case in which the $f_{esc}=1$).
The corrected color differences still show that the LyC candidates are much bluer 
than their putative LBG companions. An even much
more conservative correction can be made by assuming unitary IGM transmission
blueward of the \lya line, in all cases and recomputing the color differences 
(filled triangles), which still shows that the candidates would be much bluer 
than the LBG.

The observed median $(B-V)$ color and its 68\% percentile interval
of the LyC candidates is $0.24_{-0.22}^{+0.40}$,
significantly bluer than that of the LBG sample, which is
$1.85_{-0.58}^{+0.95}$, in contrast with what would be expected if their redshift
were the same as that of the LBG and their stellar populations were similar
(e.g., Meiksin 2006). 

We also explored a more extreme possibility by calculating at a given redshift and
$f_{esc}=1$ the expected colors arising from 
population III stars (Pop~III). The Pop~III stars of
Inoue (2011) template (with zero metallicity), convolved with the U-band filter 
and the IGM transmissions (1000 lines of sight) %and the other HST/ACS filters
are partially able to reproduce the observed $(B-V)$ colors only in the
case of high transparency of the IGM. 
Moreover, the presence of
Pop~III stars produces a very steep UV spectral slope (with a color $(i-z)=-0.2$), 
that is in contrast to what observed in our sample, 
i.e., $<(i-z)>=0.1\pm0.2$ \footnote{The bluest sources in our sample
(two with $-0.2<(i-z)<-0.1$) have a $(B-V)$ color that is bluer than the bluest 
value among the 1000 realizations calculated for Pop~III stars.}.
Therefore we exclude the presence of Pop~III stars.

Regardless of theoretical expectations, in all cases that we considered
the U--band companions have $(B-V)$ color that are bluer by more than one
magnitude (see $\Delta (B-V)$ in Tab.~1) than their corresponding LBG. If
placed at the same redshift, the line of sight to each
LyC candidate/LBG pair would be probing the IGM at transverse 
separations smaller than 20 kpc (physical) at the mean redshift of the Lyman
forest. This is much smaller than the transverse cross-correlation function
observed in QSO pairs separated of several arcminutes, for which 
the coherence length in the IGM is at least of the order of 500$h^{-1}$ kpc proper,
(e.g., Fang et al. (1996); D'Odorico et al. (1998); Rauch et al. (2005); Cappetta et al. (2010)). 
This implies that the IGM attenuation due to the forest should be
the same for the LyC candidates and their associated LBG if they were at the
same redshift. 
Since the $(i-z)$ colors of the $Uem$ sources 
are similar or even redder than those of the LBG, implying comparable or even 
redder ultraviolet spectral slope (similar stellar populations and obscuration properties),
their $(B-V)$ colors should also be similar, since the IGM attenuation
would be in practise the same.

Thus, we interpret these simulations as evidence that the LyC candidates are not regions
of the LBG where ionizing radiation is escaping from but rather relatively
unobscured star--forming galaxies at significantly lower redshift than the
LBGs.

\subsection{Photometric Redshift and SED Fitting}

To further investigate the nature of the U--band sources, namely whether
regions of the host LBG from where LyC ionizing radiation is escaping or
sources located at different, very likely lower, redshifts, we have fit the
spatially--resolved CANDELS multi--band {\it HST} (BVizJYH), ground--based (U
and K) and Spitzer/IRAC photometry to templates and spectral population
synthesis models to derive their photometric redshifts and the parameters of
their stellar populations (stellar mass, star--formation rate, dust
obscuration and age).

The images in the above bands have different angular resolution, and to
robustly measure the photometry of our sources we have used the TFIT software
package, developed by the GOODS and CANDELS teams (Laidler et al. 2007), 
which uses positional priors and PSF information to measure apparent
magnitudes in matched apertures. For each source, TFIT uses the spatial
position and morphology from the {\it HST} high--resolution images (we used
both the ACS z--band and the WFC3/IR H band) to construct a template image,
which includes nearby sources. While these are fully resolved at the {\it HST}
angular resolution, they might be marginally or even heavily blended in the
other images.  This template is then fit to the images of the object in all
other low-resolution bands after convolution with the appropriate PSF. During
the fitting procedure, the fluxes of the object in low-resolution bands are
left as free parameters. The best-fit fluxes are considered as the fluxes of
the object in low-resolution bands. These procedures can be simultaneously
done for several objects which are close enough to each other in the sky.
Experiments on both simulated and real images show that TFIT is
able to measure accurate isophotal photometry of objects to the limiting
sensitivity of the image (Laidler et al. 2007).

We derive the photometric redshift and the physical properties of the LyC
emitter candidates by fitting the observed spectral energy distributions
(SEDs) to stellar population synthesis models. Models used to measure photo-zs
are extracted from the library of PEGASE 2.0 (Fioc \& Rocca-Volmerange 1997). 
Instead of using the redshift with the minimum $\chi^2$, we integrate the probability
distribution function of redshift (zPDF) and derive the likelihood-weighted
average redshift. When the zPDF has two or more peaks, we only integrate the
main peak that has the largest power.

Since the U--band selected LyC candidates have counterparts in the {\it HST} z
and H bands in each case, we have generated two sets of fits, one based on
positional priors from the ACS z--band images, which sample the rest far--UV
SED of the galaxies at $\lambda \lesssim 1800$ \AA, and the other on priors
from the WFC3 H--band images ($\lambda\lesssim 3600$ \AA). The first (GOODS)
set uses photometry in the U and the ACS BViz bands, in the VLT/ISAAC JHK
images of GOODS-South described by Giavalisco et~al. (2004) and Retzlaff et al. (2010), 
and in the four GOODS Spitzer/IRAC bands. The second (CANDELS) set uses the WFC3/IR YJH
near--IR photometry in place of the ground--based one. The z--band positional
priors might better reflect the spatial location of the LyC emitting regions,
since essentially the same stars that power it also power most of the light in
the nLyC far UV spectrum. The z--band images also have better angular
resolution than the H--band ones. The second set offers an independent set of
measures, which takes advantage of the generally more sensitive and
higher--resolution WFC3 data; it lacks, however, the K--band data.

\begin{figure}
\epsscale{1.0}
\plotone{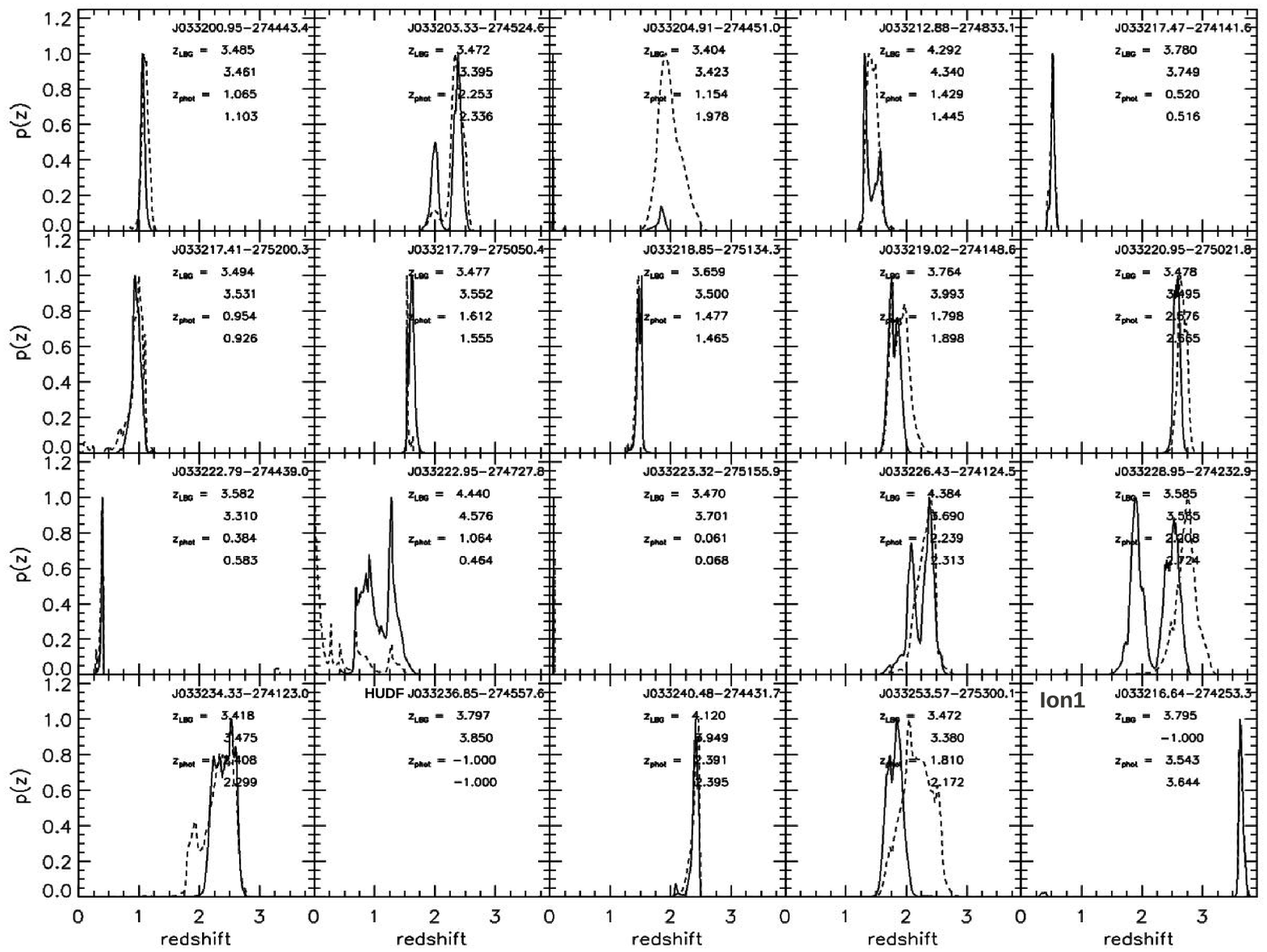} %FigGOODS2.eps} %zpdf_tb1_goods_pub2.ps}
\caption{Photometric redshift probability function of the LyC emitter
  candidates derived from the GOODS photometry in the VLT/VIMOS U, 
 {\it HST}/ACS BViz, VLT/ISAAC JHK and Spitzer/IRAC $I_1$, $I_2$, $I_3$, $I_4$
  bands in matched apertures measured with the TFIT photometric package (see
  text). The dashed curve shows the case where the U band photometry is not
  used in the calculation. The galaxy with the blank panel (and $Z_{phot}$=-1.0) 
  is the case described in Appendix.
  The two $Z_{phot}$ values (top and bottom) in each panel indicate the redshift
  calculated with and without the U-band, respectively. The $Z_{LBG}$ reports 
  the spectroscopic redshift (top) and the photometric redshift (bottom). 
  In the bottom right panel the {\em Ion1} source with LyC detection is shown.
\label{fig2}}
\end{figure}

\begin{figure}
\epsscale{1.0}
\plotone{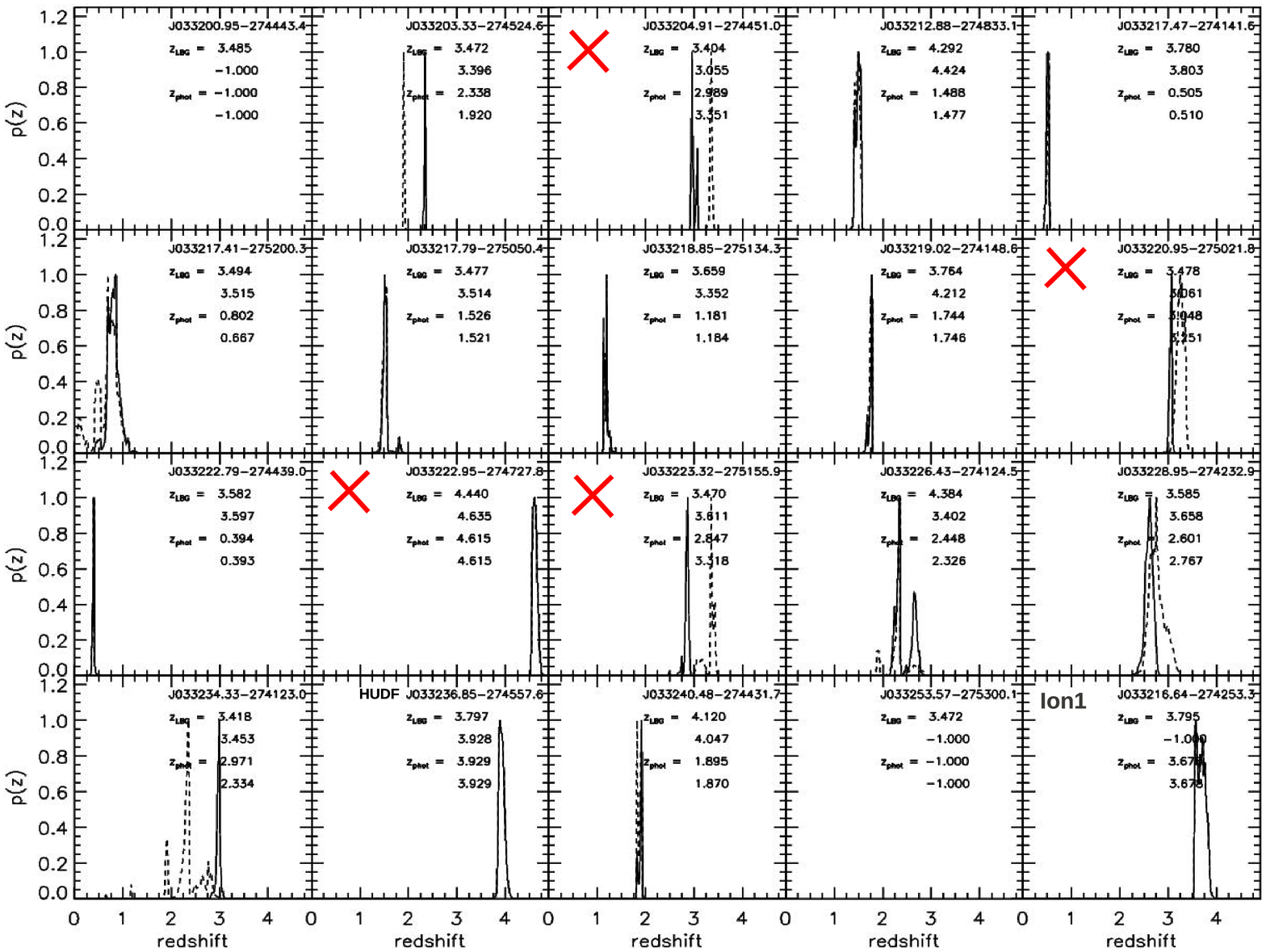} %FigCandels2.eps} %zpdf_tb1_candels_pub2.ps}
\caption{Photometric redshift probability function of the LyC emitter
  candidates derived from the CANDELS photometry in the VLT/VIMOS U, {\it
    HST}/ACS BViz, {\it HST}/WFC3 YJH and Spitzer/IRAC $I_1$, $I_2$, $I_3$,
  $I_4$ bands in matched apertures measured with the TFIT photometric package
  (see text). The dashed curve shows the case where the U band photometry is
  not used in the calculation. A galaxy with a blank panel (and $Z_{phot}$=-1.0)
  means that no {\it WFC3} data are available for that source. 
  The two $Z_{phot}$ values (top and bottom) ineach panel indicate theredshift
  calculated with and without the U-band, respectively. The $Z_{LBG}$ reports
  the spectroscopic redshift (top) and the photometric redshift (bottom).
  In the bottom right panel the {\em Ion1} source with LyC detection is shown.
  Crosses are the $Uem$ sources for which the photometric redshift is not useful 
  (see text for further details).
\label{fig3}}
\end{figure}

Since the potential emission in the LyC spectral range is not taken into
account in the models, and thus flux in the U--band could skew the
photometric redshift calculation towards lower values, for both photometric
sets we have run the fit with and without the U--band photometry. Figure~\ref{fig2}
shows the redshift probability function derived from the GOODS
photometric set; Figure~\ref{fig3} the one from the CANDELS set. In all cases
the solid curve refers to the calculation made using the U--band photometry,
the dashed curve without it. In all cases the photometric redshifts are
calculated as the weighted average of zPDF as:

\begin{equation} 
z_{phot} = {\int_0^\infty z\, P(z)\, dz \over \int_0^\infty P(z)\, dz }.
\end{equation}

\noindent A blank panel in Figure~\ref{fig3} means that the source is outside 
of the region covered by the CANDELS observations.

The general result from our analysis is that the photometric redshifts of the
LyC emitter candidates are systematically much smaller than the spectroscopic
redshifts of their host LBGs. Exceptions include the cases of the LyC
candidates GDS~J033204.91-274451.0, GDS~J033220.95-275021.8,
GDS~J033222.95-274727.8, and GDS~J033223.32-275155.9 (see Figures~\ref{fig2} 
and ~\ref{fig3}), where the GOODS photometric redshift 
is much lower than the spectroscopic redshift of the LBG, but the CANDELS photometric 
redshift is not. These are cases in which the separation between the LBG and the $Uem$
is $\lesssim$1\see~and are not individually detected 
in the CANDELS H-band based catalog (they are marked with red crosses in Figure~\ref{fig3}).

$\bullet$ In the first case, GDS~J033204.91-274451.0, the LyC emitter candidate is resolved
as a different source from the LBG in the ACS z--band image but not in the
H--band one. The photometric redshift from the GOODS photometry yields a
significantly lower redshift than the LBG's spectroscopic one, with some
dependence of the redshift probability function on the exclusion of the 
U--band from the fit, but the resulting photometric redshift is in any case
much lower than the LBG's. Since the LyC candidate and the LBG are not
resolved in the H band, the CANDELS photometric redshift is essentially that
of the LBG itself. If the U--band photometry is not included in the fit, the
spectroscopic and photometric redshift are in very good agreement. Including,
the U--band data, however, yields a photometric redshift somewhat lower than
the spectroscopic one ($z_{spec}=3.404$,$z_{phot}=3.351$), which is not
surprising, since this is the LBG with the lowest redshift in our sample and
thus some flux detected in the bluest band might easily skew the fit toward
lower redshift values.

$\bullet$ The other three sources (GDS~J033222.95-274727.8,  GDS~J033220.95-275021.8,
GDS~J033223.32-275155.9) are also cases in which the LyC candidate and the LBG 
are classified as individual
sources in the z--band but not in the H band. In both cases, the GOODS
photometric redshift is much lower than the LBG's spectroscopic one. In the
former, since the LBG redshift is at the high end of our sample
($z_{spec}=4.440$), adding or not the U--band data to the derivation of the
photometric redshift makes no difference. In the other two, 
neglecting the U--band data results in
very good agreement between the spectroscopic and photometric data, while
including it lowers the photometric redshift solution.

The LyC emitter candidate with coordinates {\it RA=03:32:36.85}, {\it DEC=-27:45:57.6},
associated with galaxy HUDF-J033236.83-274558.0 is discussed in detail in the Appendix. 
It is the fainter among the cases discussed in this work with an U mag of 28.6. 
A new observed feature in the J-band (at 10$\sigma$ significance) is consistent with 
relatively strong line emission of a low redshift galaxy, lower than its companion LBG.

A general characteristic of the fits is that including or not the U--band
photometry in the derivation of the photometric redshift of the LyC emitter
candidates makes no substantial difference in the results and that there is very good
quantitative agreement between the GOODS and CANDELS photometric redshift,
pointing to the robustness of our analysis. 

Finally, a further check has been performed by calculating the photometric
redshifts without the inclusion of the U and $B_{435}$ bands, that makes the test
independent from the single $B-V$ color analysis described in Sect. 4.2.
Even though the efficiency of the photometric redshift prediction is slightly reduced,
the resulting $z_{phot}$ best-fit values still favor low redshift solutions for all sources, 
except for GDS~J033223.32-275155.9 that increases from $z_{phot}=0.07\pm0.02$ to 
$3.60 \pm 0.2$. In this case, however, the $(B-V)=0.65$ color, the relative escape fraction 
and the $\Delta (B-V)$ measurements are in contrast with this high redshift solution.

\subsection{Conclusion From $f_{esc}$, UV Colors and Photometric Redshift Analysis}
In summary, the combination of photometric redshift, relative UV colors 
(that trace the IGM decrement) and the relative escape fraction estimated for 
the LyC candidates all argue against them being ionizing radiation escaping form the nearby
LBG. Rather, they are most likely low--redshift interlopers (as we had
originally argued in V10b). The more extreme case among the sources here 
analyzed (HUDF~J033236.83-274558.0) is discussed in detail in the
Appendix, in which we report new observational evidence supporting 
a possible contamination from a superimposed object.

Finally, we conclude by noting that deep UV imaging with {\it HST}/WFC3
will be able to clarify the nature of the U--band offset LyC candidates, since
the probability that LyC radiation at $\lambda \ll 912$\AA\ escapes the host
galaxy and eludes the IGM is extremely low (for an estimate of this
probability at $\lambda <700$\AA\ see Inoue \& Iwata 2008; Meiksin 2006).

\section{A Critical Analysis Of The Current Measurements}

In this section we review all known (to us) galaxies at high redshift 
($z\gtrsim$3) for which emission of LyC ionizing radiation has been reported or suspected.

$\bullet$ Direct detection of LyC ionizing radiation has been reported by Shapley et
al. (2006, S06) in deep spectra of two LBGs at $z\sim 3$ (U--band
dropouts) in the SSA22 field, dubbed D3 and C49 in their nomenclature, out of
a total of 14 sources observed with comparable sensitivity. 
It is worth noting that I09 and N11 found a highly significant null detection
for the object SSA22a-D3, which is the brighter of the two
objects for which S06 claimed an ionizing flux detection.

$\bullet$ Iwata et~al. (2009) used deep narrow--band (NB)
imaging to image the rest--frame SED blueward of the 912\AA\ Lyman limit of
members of a well--known overdensity of galaxies at $z\sim 3.1$, also in the
SSA22 field (Steidel et~al. 1998). They reported 17 detections, 7 from galaxies
selected as LBGs and 10 as LAEs. 

$\bullet$ Nestor et~al. (2011) used the same technique to observe
galaxies in the same SSA22 field, but extended the sensitivity of previous
observations by $\approx 0.6$ mag. They reported 34 detections, out of 156 
sources, which include 6 LBGs and 28 LAEs.

$\bullet$ In the paper V10b, we used a technique similar, but not identical, to the one
by I09 and N11. Instead of using a narrow--band, we carried out ultra--deep U--band
imaging to search for candidate LyC emitters among galaxies at $3.4\lesssim
z\lesssim 4.5$, namely such that the redshifted $912$\AA\ ionization edge of
the targeted sources is to the red and completely outside of the filter's
bandpass. 
We found 1 candidate out of 102 LBGs (described in Sect. 6).

For most of the NB--selected candidates and LBGs the region where the LyC ionization
radiation originates is found to be spatially offset from that of the non-ionizing
far--UV light, namely the main body of the galaxy. The displacement is
generally less than one arcsec, but in some cases values as high as 2$''$ have
been reported. We performed a visual
inspection of the images of the galaxies by I09 and N11, and consulted Table~4
and 5 of N11, and found that $\gtrsim 70$\% of the presently identified LyC
emitter candidates exhibit a spatial offset between the LyC emission and the
``non-ionizing'' far--UV image of the galaxies.
Since an offset emission is also consistent with a lower redshift
interloper, a key step is to measure the statistical occurrence of these cases
(as we did in V10a).
In particular with availability of large--area {\it HST} UV and near--IR surveys with
WFC3, such as CANDELS, the prospect for substantial progress in this regard is
good. The other important point is that if the offset LyC emission is real,
then care must be exercised when calculating the fraction of escaping ionizing
radiation $f_\mathrm{esc}$ in each case, i.e., whether it is from the galaxy as
a whole or only from a close companion, to avoid bias in the derivation of
average properties. For example it is worth mentioning that the 
offset possible LyC detection in S06, SSA22-C49, has a reported observed ratio 
of $(F1500/F_{LyC})_{OBS} = 16.4 \pm 6.1$ (N11),
in which the flux $F1500$ used in the calculation is that of the main galaxy.
Presumably, a calculation restricted to the local region where the putative 
Lyman continuum emission is seen would produces a much smaller value, i.e.,
a much higher $f_{esc,rel}$.

Finally, it must be kept in mind that part of the candidates LyC emitters
reported above may have the contribution to $f_\mathrm{esc}$ from an AGN
component that is difficult to identify without multi-wavelength and
spectroscopic surveys.

Therefore the amount of $stellar$ LyC emission from high-z star-forming galaxies
is still uncertain. Very few sources have spectroscopic redshift confirmation:
(A) those (five) reported by Inoue et al. (2011) that  need 
extreme stellar populations to justify the flux ratios observed,
(B) the LBG reported in S06 (SSA22-C49) that shows an offset (0.5\see) U-band emission
whose redshift is not conclusively known, and finally (C) the one we present in the 
next section and discovered in the GOODS-S field ({\em Ion1}). 

\section{A LyC Emitter Found in the GOODS sample: {\em Ion1}}
The direct detection of LyC emitters at high redshift (as high as possible),
allow us to characterize their properties at $\lambda >1215.7$\AA~and try to
identify similar sources and/or look if they are more common during the
re--ionization epoch.  For this reason it is important to identify secure LyC
emissions.  We recall one of the most promising stellar ionizers at
redshift 3.795 we have identified in the GOODS-S field and currently the
highest redshift known so far (GDS~J033216.64-274253.3, {\em Ion1}) and
is useful to revisit it in the context of this work by comparing its appearance
with the offset $Uem$ sources discussed above.

\subsection{Observed Photometric and Spectroscopic Properties of {\em Ion1}}

This source was initially reported in V10b, and here we briefly summarize and 
report on new observational constraints, 
and compare it with the offset $Uem$ sources discussed above.

\begin{figure}
\epsscale{1.0}
\plotone{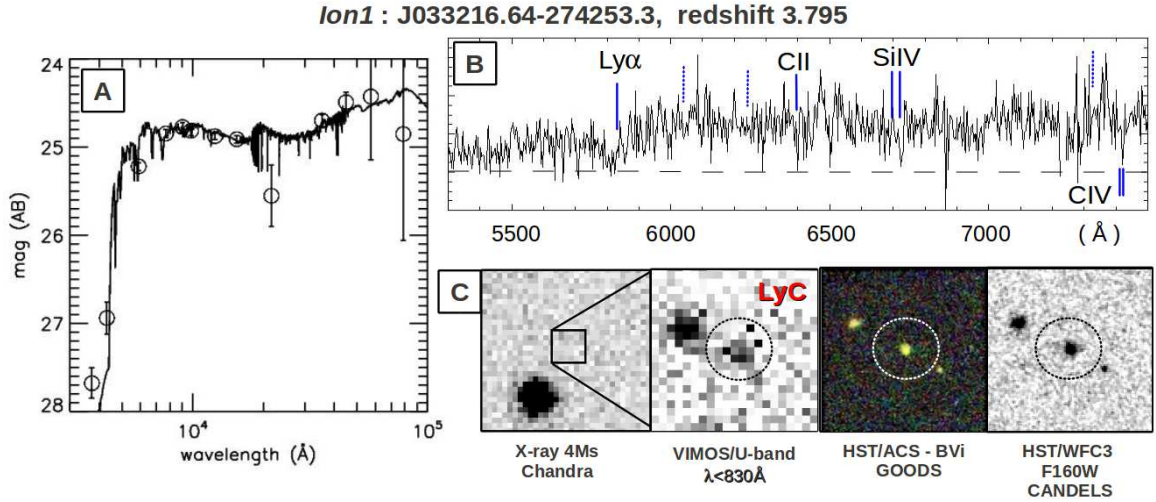} %f3.ps} %test.ps} %f3.eps}
\caption{SED fitting (left, panel A), Keck/DEIMOS spectrum (top right, panel B) and image cutouts of the 
{\em Ion1} galaxy (bottom right, panel C) are shown. 
{\bf Panel A:} The SED fitting is shown and includes from left to right the VLT/VIMOS $U$--band,
{\it HST}/ACS $B_{435}$, $V_{606}$, $i_{775}$, $z_{850}$ bands and the {\it HST}/WFC3 $Y$, $J$ and $H$ bands,
the VLT/ISAAC $K_{S}$--band (the most deviant point), and the four $Spitzer$/IRAC channels
3.6$\mu m$, 4.5$\mu m$ 5.8$\mu m$ and 8.0$\mu m$ (only two points come from ground-based observations).
{\bf Panel B:} Keck/DEIMOS spectrum with UV absorption features indicated that yield a redshift $z=3.795$ 
(solid vertical lines and labels). Dotted vertical lines
mark the expected positions of [O\,{\sc i}]1302.2, [Si\,{\sc ii}]1260.4 and [Si\,{\sc ii}]1526.7,
from left to right, respectively. {\bf Panel C:} Image cutouts from the X-ray to the $U$--band
rest-frame are shown (the VIMOS, ACS and WFC3 box sizes and the square inset in the X-ray image 
are 6.3\see~on a side). Note that the VLT/VIMOS $U$-band is sampling the wavelengths
smaller than 830\AA~rest-frame.
\label{fig6}}
\end{figure}

{\bf SED:} The source has not been detected in the F225W, F275W and F336W
channels of {\it HST}/WFC3 observations in the GOODS-S (down to 26.3, 26.4, 26.1 at
5$\sigma$ for point-like objects, Windhorst et al. (2011)).  It lies
in the B-band dropout selection scheme, with (B-V)=1.70 and (V-z)=0.40 (V09),
and has a blue UV spectral slope, having an $(i-z)=-0.015$, and a $\beta=-2.09 \pm 0.16$
derived from Castellano et al. (2011).
It is worth noting that this source has the bluest $(B-V)$ color among the
available 17 LBGs with $3.7<z<3.9$ and comparable UV slope. 
This is consistent with the fact that the $B_{435}$--band is probing the
rest-frame interval $\lambda \lesssim 1020$\AA, in which the LyC
is contributing to the observed flux (a S/N of 10 is measured in this band) and
reduces the $(B-V)$ color.
The probability that $zphot>3.4$ is $\simeq$ 100\%.  It has not been detected
in the new 4 Ms Chandra X-ray observations, for which we set a 1$\sigma$ upper limit
of $L_{X} < 3 \times 10^{42} erg/s$, nor in the 24$\mu$m SPITZER/MIPS observations
(Santini et al. 2009). The SED is fully comparable to that of a
star-forming galaxy, in particular from the SED fitting we obtain : $SFR
\simeq 50 M_{\odot}yr^{-1}$, stellar mass of $2.3\times10^{9}M_{\odot}$,
$E(B-V) \lesssim 0.1$ (see Figure~\ref{fig6}).

{\bf Optical spectrum:} Optical spectra from Keck/DEIMOS and VLT/VIMOS have
been obtained, although the latter is too shallow and the S/N prevent us to add
information.  From the Keck spectrum it is clear that the \lya line is not in
emission, the continuum--break is evident together with the [Si\,{\sc
    iv}]1393.8--1402.8 and [C\,{\sc iv}]1548.2--1550.8 absorption lines, at
redshift 3.795.  Also [C\,{\sc ii}]1335.1 in absorption seems to be
present. The other UV absorption lines like [O\,{\sc i}]1302.2, [Si\,{\sc
    ii}]1260.4 and [Si\,{\sc ii}]1526.7 are not detected (see Figure~\ref{fig6}).

{\bf UV and B-band rest-frame morphology:} At 1700\AA/1900\AA~and B--band
rest-frame wavelengths the galaxy shows a resolved and compact shape, with an
half-light radius of 0.9 kpc (physical size) in both the $z_{850}$ and $H$ bands (the latter
from CANDELS), respectively (see Figure~\ref{fig6}).

The hypothesis that it is a faint AGN ($L_{X}< 3 \times 10^{42} erg/s$) is
still open. However, the spectral features in the
optical spectrum and the shape of the SED from the U--band to the far infrared
(Spitzer/MIPS 24$\micron$) are fully compatible with
a star-forming and relatively low mass galaxy. In the following we assume that
it is a star-forming galaxy. 

The LyC and the nLyC emissions
(namely U and $i_{775}$--bands) are spatially aligned within the errors,
$\Delta \theta < 0.1$\see, and the LyC detection at $\lambda<830$\AA~implies a
relatively transparent line of sight free from LLSs.  From Eq.~\ref{eq:f_esc1} we derive a
minimum $f_{esc,rel}=82\%$ assuming $(L1500/L_{LyC})_{\rm int}=7$ and a IGM
transmission of 0.575 (maximum value for our U-band filter of the 10000 
realizations at z=3.8).
Given its $E(B-V) \lesssim 0.1$ and adopting the Calzetti extinction law we have
$f_{esc}>56\%$ (Calzetti et al. 2000). Assuming a intrinsic value for $(L1500/L_{LyC})_{\rm int}=3$
the escape fractions are $f_{esc,rel}>35\%$ and $f_{esc}>24\%$.

\subsection{How Does {\em Ion1} Compare With the Rest of the LBG Population ?}

It has been firmly established that as the
\lya equivalent width increases from absorption to emission, the UV spectral
slope becomes bluer and the strength of the UV absorption lines (stellar and
interstellar) drastically decreases (see, e.g., Shapley et al. 2003; Kornei et
al. 2010; Balestra et al. 2010; Vanzella et al. 2009). Moreover, on average,
high \lya equivalent widths tend to be associated to UV compact morphologies;
conversely sources with \lya in absorption appears less compact (see Law et
al. 2007; V09; Pentericci et al. 2010).  The {\em Ion1} galaxy shows :
\begin{enumerate}
\item{a blue UV spectral slope $\beta=-2.09 \pm 0.16$, 
from the photometric fitting of Castellano et al. (2011).}
\item{relatively weak UV absorption lines, from the Keck/DEIMOS spectrum
(see Figure~\ref{fig6}).}
\item{a compact morphology, 0.9 kpc physical half light radius in the 
1600\AA~and 3300\AA~rest-frame wavelengths (from HST/ACS and WFC3).} 
\end{enumerate}

These are all features that positively correlate with the equivalent width
of the \lya in emission, nonetheless the \lya is absent.  In this respect it deviates
from the average LBG population, particularly if we consider its LyC emission.
The fact that the \lya line is not in emission suggests that it can be absorbed
and/or it is intrinsically weak because the ionizing radiation is escaping the
galaxy. 
The detection of LyC emission would suggest low gas attenuation along the line of
sight and the blue UV slope low dust absorption. 
In the extreme case in which all the ionizing radiation is escaping
($f_\mathrm{esc}$ is 100\%, or is close to), nebular emission lines like \lya,
Oxygen or the Balmer transitions are no longer pumped, and therefore drop to
intrinsecally small equivalent widths (Schaerer 2003). We can image a geometrical
configuration in which there is a high fraction of escaping ionizing
radiation and at the same time a high \lya~equivalent width in emission; it
can be realized with a sort of ``unipolar'' outflow behind the source that
backscatters the \lya~photons to the observer (as is usually seen), while along
the line of sight (the front of the galaxy), the ISM and CGM are free from
dust or gas attenuation and the LyC radiation can escapes.  However,
situations like this have never been clearly observed until now and the
interplay between the escape fraction of ionizing and \lya photons is still
not known.

It is worth noting that the absence of \lya emission from an object with
high $f_\mathrm{esc}$, like {\em Ion1}, makes extremely difficult the
spectroscopic confirmation of similar sources at $z>7$, i.e. during the reionization epoch.
Indeed, if place at $z=7$ it would be one magnitude fainter, $Y \simeq 25.9$,
and the continuum break hard to measure spectroscopically 
(Vanzella et al. 2011; Fontana et al. 2010; Pentericci et al. 2011).

\section{Conclusions}

The observation of the LyC in distant galaxies is a difficult task because
several attenuations occur: in the galaxy itself and circumgalactic medium
(by dust and nutral hydrogen gas) and along the intergalactic travel 
(Lyman alpha forest, LLSs and DLAs). 
A further complication is related to the possible presence of 
foreground (lower-z) superimposed sources that can mimic the LyC emission 
of the background galaxy.
It is therefore crucial to establish whether they are genuinely associated
with the redshift of the main LBG.

In this work we have discussed three diagnostics that can be used 
to constrain the redshift of these offset and faint sources.
To this end, the deep and high angular resolution multi-wavelength images 
available from the GOODS and CANDELS surveys have been exploited.
In particular, starting from the sample of U-band emitters identified
in V10b, we (1) calculated accurate photometric redshifts, (2) considered the
IGM radial and transverse absorption (simulating UV colors and comparing
those of LBGs and their U-band emitters) and (3) provided new constraints
on their nature by calculating the $f_{esc}$ quantity under conservative assumption.
These three analysis suggest that none of offset U-band sources are 
actual sources of LyC emission from the LBGs at $z>3.4$,
but are instead foreground interlopers.
This strengthens the results of V10b, i.e., 
the median $f_{esc}$ quantity is very small for $L>L^{*}$ galaxies, 
or assuming a bimodal distribution for the $f_{esc}$, the
higher values are rare.

At present, the offset candidates reported in literature by S06, I09 and N11 need 
spectroscopic redshift confirmation and are difficult to access if deep 
multi-wavelength high angular resolution imaging is not available.
%More effort is needed to try to confirm the redshifts of 
%these offset emitters.

Moreover, we note that in order to avoid biased measures,
the calculation of the $f_{esc}$ quantity in the case of spatially 
offset emission must be performed carefully considering 
quantities arising from the same physical region.
We argue that the measurements reported by S06, I09 and N11 are
affected by this problem and need futher investigations.

The current situation is far from clear. The number of {\it bona fide} LyC
detections at $z>3$ is very small and not statistically significant.

We have discovered one good candidate (named {\em Ion1}) that is currently the 
highest redshift galaxy known with direct LyC detection.
Its LyC emission at $\lambda <830$\AA~(probed by the U-band)
is aligned with the source detected in the UV nLyC, i.e., no offset
is observed with this resolution.
Apart from the three cases reported by Inoue et al. (2011) 
with aligned LyC emission that need Population-III stars to explain
their observed $f1500/f_{LyC}$ flux ratios (even smaller than one), 
the {\em Ion1} emission is easily explainable with standard stellar 
population and an $f_{esc}>25\%$.
If the AGN component is negligible, as seems to be the case from the X-ray and
spectroscopic data, it would be the most promising stellar ionizer
that may resemble those responsible for HI reionization.

It is worth noting that a high value of the escape fraction 
would correspond to faint nebular emission (e.g., Robertson et al. 2010). 
In this case the \lya\ line is not in emission, despite the fact that the source shows a 
blue UV spectral slope ($\beta=-2.09$), relatively weak interstellar
absorption lines and compact morphology, all characteristics that 
positively correlate with the strength of the \lya\ emission line
(e.g., Shapley et al. 2003; V09; Pentericci et al. 2010; Balestra et al. 2010).
It is not clear if the absence of \lya\ is connected to the high value of 
the $f_{esc}$ or due to other effects like the dust attenuation, that
however, would be small, giving the blue UV slope.

Its relatively low stellar mass ($2.3\times10^{9}M_{\odot}$) may also 
be a signature that feedback processes were more efficient to clean the line of sight 
in this type of galaxy, favoring the escape of ionizing radiation. Precisely in this regard 
a measure of the wind velocity is important (Heckman et al. 2011; Overzier et al. 2010).
We note that {\em Ion1} is not a LAE,
therefore it is important to complement the LAE surveys with sources
like this. Clearly sources like {\em Ion1} placed at z$>$7 would be extremely 
difficult to detect spectroscopically with present facilities.
However, since at $z>7$ the direct measure of LyC is not feasible, 
sources like {\em Ion1} are the only viable ways we have to investigate the 
mechanism that allow the ionization ratiation to escape, i.e.,
to address the interplay between the $f_{esc}$ quantity and the non-ionizing UV features
like the strength of the nubular emission lines (e.g., Oxigen, Balmer and Lyman lines),
UV slopes, nature of the stellar populations, geometry, winds, etc.

\acknowledgments 
%Observations were carried out using the Very Large
%Telescope at the ESO Paranal Observatory under Programme IDs
%085.A-0844, 283.A-5052 and 181.A-0717. 
%{\bf We thank the anonymous referee for useful comments.} 
%We would like to thank P. Dayal and A. Ferrara for useful comments and 
%discussion. 
We acknowledge financial contribution from the agreement ``COFIS'' ASI-INAF 1/009/10/0.
We would like to thank A. K. Inoue for providing us the transmissions of the
IGM and the spectral template of Pop~III stars.

\clearpage

\clearpage

\clearpage

\clearpage

\clearpage

\clearpage

\clearpage

\clearpage
\begin{table}
%\scriptsize                                                                                                                                            
\begin{center}
\caption{Sample of offset U--band emission from LBG in GOODS-S. \label{tab1}}
\begin{tabular}{lcll|ccc|cc}
\tableline\tableline
 GOODS-ID         &  $\Delta\theta$ & $zspec$  &$zphot^{c}$~($Uem$)& $S/N_{B}$ & $\Delta (B-V)_{obs}$ &  $\Delta (V-i)_{obs}$ &$\left( \frac{f1500}{f_{LyC}}\right)_{OBS} $ & $f_{esc,rel}^{a}$\\
 ($Uem$)          &  (arcsec)       &  LBG     & GOODS,CANDELS& ($Uem,LBG$)  & (LBG-$Uem$)          &  (LBG-$Uem$)         &                                           &  (\%)\\
\tableline
 J033200.95-274443.4           &      1.45 &  3.485&1.10(0.06),-1.00  & 23,27  &1.06  & -- & 1.80&251\\ %corr 0.11 a z=3.5   %U(V-z)=0.78  LBG(V-z)=0.0448 (*)  Ui-z=0.3005 
 J033203.33-274524.6           &      1.90 &  3.472&2.34(0.15),1.92(0.08)&  97,26  &1.13  & -- & 1.20&376\\    %U(V-z)=-0.14 LBG(V-z)=0.21        Ui-z=-0.0588    
 J033204.91-274451.0$^{\star}$  &      0.70 &  3.404&1.98(0.23),-1.00&  19,16  &1.71  & -- & 0.66&792\\ %corr 0.07 a z=3.4  %U(V-z)=-0.47  LBG(V-z)=0.52 (*)   Ui-z=-0.176
 J033212.88-274833.1           &      1.50 & 4.292 &1.44(0.09),1.48(0.05)&  30,$<$1  &--    & 1.15 & 1.01&724\\   %U(V-z)=0.32  LBG(V-z)=1.23        Ui-z=0.328    V-i=-7.3013e-03      
 J033217.47-274141.6           &      1.20 &  3.780&0.52(0.04),0.51(0.03)&  20,4  &1.00  & -- & 6.85&76\\ %corr 0.36 a z=3.8  %U(V-z)=0.82  LBG(V-z)=0.67 (*)     Ui-z=0.1766   
 J033217.41-275200.3           &      1.40 &  3.494&0.93(0.18),0.67(0.24)&  10,15  &1.07  & -- & 2.29&197\\  %U(V-z)=0.70  LBG(V-z)=-0.09 (*)    Ui-z=0.01401                     
 J033217.79-275050.4           &      1.60 &  3.477&1.55(0.03),1.52(0.08)&  8,13   &1.58  & -- & 1.77&255\\   %U(V-z)=0.69  LBG(V-z)=0.69 (*)    Ui-z=0.4364                         
 J033218.85-275134.3           &      1.30 & 3.659 &1.46(0.06),1.18(0.05)&  35,4  &2.17  & -- & 1.31&373\\ %corr 0.245 a z=3.7  %U(V-z)=0.27  LBG(V-z)=0.62    Ui-z=0.281            
 J033219.02-274148.6           &      1.10 &  3.764&1.90(0.14),1.75(0.03)&  20,$<$1  & 2.82 & -- & 1.54&339\\   %U(V-z)=0.09  LBG(V-z)=1.02        Ui-z=0.0242                
 J033220.95-275021.8$^{\star}$  &      0.60 &  3.478&2.66(0.07),-1.00&  14,6  & 1.22 & -- & 3.08&146\\     %U(V-z)=0.03  LBG(V-z)=0.44       Ui-z=0.0123              
 J033222.79-274439.0            &     1.30 &  3.582&0.58(0.75),0.39(0.02)&  28,11 &0.85   & -- & 2.27&210\\ %corr 0.17 a z=3.6  %U(V-z)=0.14  LBG(V-z)=0.20 (*)  Ui-z=0.118         
 J033222.95-274727.8$^{\star}$  &      0.80 &  4.440&0.46(0.46),-1.00& 4,$<$1  &--  & 1.05 & 2.31&325\\      %U(V-z)=1.07  LBG(V-z)=1.71  Ui-z=0.4913  V-i=5.8250e-01, V-i(LBG)=1.63
 J033223.32-275155.9            &     1.10 & 3.470&0.07(0.07),-1.00& 19,12 &1.36  & -- & 4.45&102\\     %U(V-z)=0.24  LBG(V-z)=0.75      Ui-z=0.0427             
 %J033225.09-274852.6$^{\star,c}$  &      0.70 & $<$0.1\%  & 3.484&  1.84 &9 &-0.26 &  0.0  &  0.0& 1.6$\pm$0.1\\         
 J033226.43-274124.5$^{\star}$  &      0.70 &  4.384&2.31(0.16),2.32(0.20)&  19,4 &--  & 0.56 & 2.68&280\\       %U(V-z)=0.00  LBG(V-z)=0.61       Ui-z=0.0482  V-i=-4.5300e-02 
 J033228.95-274232.9           &      1.40 &  3.585&2.72(0.18),2.77(0.17)&  12,11 &1.05  & -- & 2.42&197\\       %U(V-z)=0.08  LBG(V-z)=0.21 (*)   Ui-z=0.0409                      
 J033234.33-274123.0           &      1.60 &  3.418&2.30(0.25),2.33(0.31)& 17,11 &1.31  & -- & 2.86&148\\     %U(V-z)=-0.06  LBG(V-z)=0.55        Ui-z=-0.122                        
 J033236.85-274557.6$^{b}$     &      0.40 &  3.797&-1.00,-1.00  &  5,6 &1.80 & -- & 1.36&387\\       %U(V-z)=0.0   LBG(V-z)=0.61        Ui-z=                          
 J033240.48-274431.7           &      1.40 &  4.120&2.39(0.07),1.87(0.04) & 36,2 &--   & 0.54 & 1.24&505\\       %U(V-z)=-0.13  LBG(V-z)=0.48      Ui-z=-0.0838  V-i=-4.7199e-02     
 J033253.57-275300.1           &      1.10 &  3.472&2.17(0.25),-1.00  & 9,16  &1.06   & -- & 1.66&272\\      %U(V-z)=-0.18   LBG(V-z)=0.25      Ui-z=-0.218
\hline
 ({\em Ion1})                  &$<$0.1\see &  3.795&3.64(0.04),3.68(0.09)  & 10.3  &--   & -- & 15.00 &35\\
\tableline
% J033216.64-274253.3$^{d}$     &    $<$0.1 & 100\%& 3.796&  1.70 & 10 & 0.0 & -- & --\\ 
%1  &  (1.84),(1.92)        & 0.7\see   & 26.10 & 26.79 (-0.18)- 0.0 & 12.1, $2\%$,$<<1\%$ & 04.87-274451.4 (3.404,A)\\ %04.91-274451.0                
%2  &  (0.06),(0.07)        & 1.1\see   & 26.29 & 24.67 (0.04)- 0.81 & 3.4, $42\%$,$<<1\%$ & 23.34-275156.9 (3.470,A)\\ %23.32-275155.9                
%3  &  (1.51),(1.46)        & 1.3\see   & 25.17 & 24.88 (0.28)- 3.33 & 2.2, $<<1\%$,$22.5\%$ & 18.83-275135.5 (3.659,A)\\ %18.85-275134.3              
%4  &  (2.62),(2.64)        & 0.6\see   & 27.16 & 25.94 (0.01)- 0.47 & 6.6, $29\%$,$<<1\%$ & 20.97-275022.3 (3.478,A)\\  %20.95-275021.8               
%%26.43-274124.5 &  zphot   & 0.9\see        & 26.80 & 25.73    & 0.28 & 26.49-274124.0 (4.384,B)\\                                                    
%%22.95-274727.8 &  zphot   & 1.2\see        & 27.58 & 26.66    & 0.32 & 22.88-274727.6 (4.440,B)\\                                                    
%\tableline                                                                                                                                            
%5 $^{b}$       &  $z\simeq2.5$   & 0.4\see & 28.63 & 28.30 (0.00)- 0.37 & 45.2, $<<1\%$ & 36.83-274558.0 (3.797,A)\\ % see yicheng@astro.umass.edu    
%\tableline                                                                                                                                            
%               &  $-$   & 0.0\see        & 27.79  & 24.86 (-0.02)- 0.21 & $30\%$,$4.5\%$ & 16.64-274253.3 (3.795,A)\\                                 
\tableline
\end{tabular}
\tablecomments{(a) These are the {\it minimum} $f_{esc,rel}$ derived from Eq.~\ref{eq:f_esc1} adopting
conservatively the maximum IGM transmission at the given redshift. 10000 IGM transmissions have been calculated
from simulations specifically computed at the spectroscopic redshift of the LBG.
The intrinsic ratio has been assumed to be $L1500/L_{LyC}=3$ (see text for details).
(b) galaxy observed in the HUDF and shown in Figure~\ref{fig7}. The angular separation
between the compact blue emitter and the center of light of the LBG is 0.4\see. (c) The photometric
redshifts with their r.m.s. (GOODS and CANDELS, on left and right, respectively) without the inclusion 
of the U-band are reported. ($\star$) sources shown in Figure~\ref{fig1}.}
%(c) This is the estimated $A_{V}$ from the SED fitting fixing the                                                                                     
%redshift of the $Uem$ to that of the LBG (column 4) and metallicity to the Solar.}                                                                    
%(c) Despite the high B-V color, this is an evident foreground red galaxy, see Fig.~1.                                                                 
\end{center}
\end{table}

\appendix
\section{The Case Of The LBG HUDF~J033236.83-274558.0}

\begin{figure}
\epsscale{1.0}
\plotone{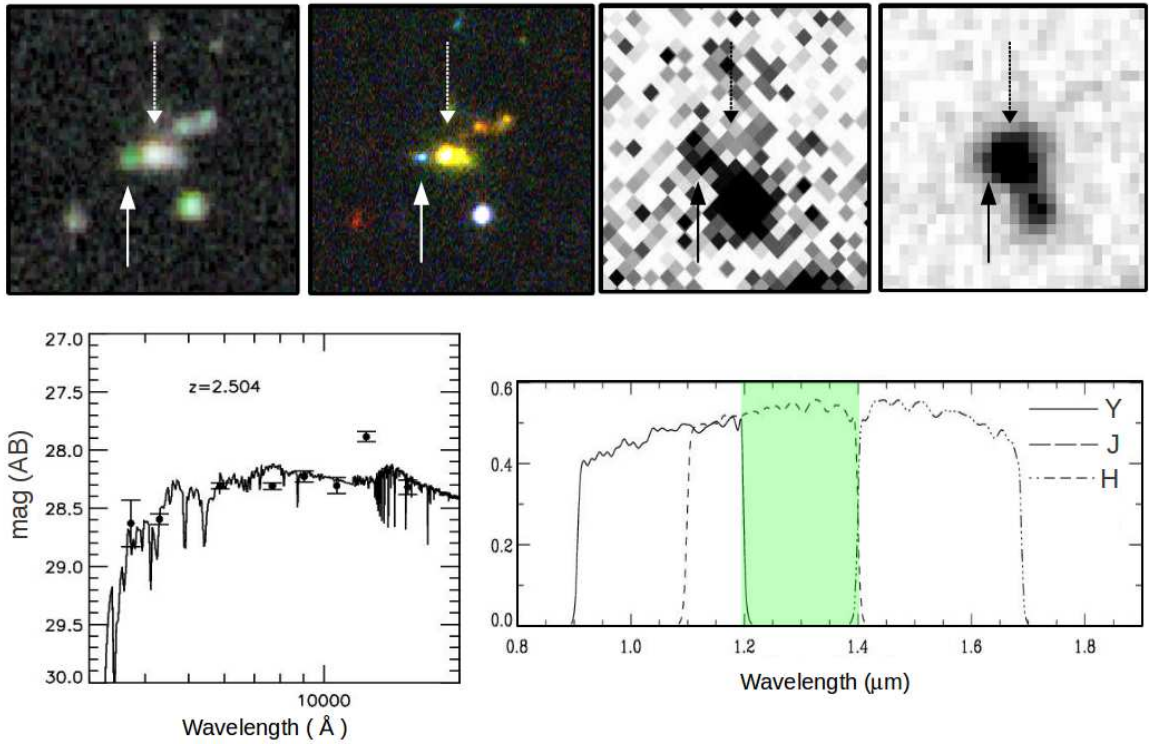} %f2.eps}
\caption{Images and spectral energy distribution for J033236.83-274558.0 in the HUDF.
In the top part, from left to right: the deep color composite obtained with {\it HST}/WFC3
F105 ($Y$), F125 ($J$), F160 ($H$) bands, the BVI color {\it HST}/ACS composite, the ultra-deep
VLT/VIMOS U--band image (that is probing the LyC for the LBG) and the deep VLT/VIMOS $R$ band image.
The box sizes are 4.5\see~on a side. 
Note that such a system observed from the ground is poorly resolved ($R$--band on the right).
The dotted arrows indicate the position of the
LBG at $z=3.797$ and the solid arrows mark the position of the blue emitter.
The angular separation between the blue emitter and the peak of the LBG emission is
 $\simeq$ 0.4\see. However the LBG extends beyond this separation and the blue spot is fully superimposed.
This compact blue source shows also a boosted emission in the $J$ band,
$\sim$ 0.4 magnitudes brighter (10$\sigma$) than the other two $Y$ and $H$ (bottom left),
it is the reason why it appears as a green spot in the WFC3 composite. This 
$J$--band emission is compatible with a low redshift interpretation
($zphot < 3$, see text for details). In the bottom right, the {\it HST}/WFC3 relative transmissions of the
$Y$, $J$ and $H$ filters are shown. The shaded green region mark the wavelength range where
the presence of the emission line(s) are boosting the $J$--band flux. 
 \label{fig7}}
\end{figure}

Figure~\ref{fig7} shows the {\it HST}/ACS images of the offset U--band LyC emitter
candidate that we have identified at $0.4$\see\ from LBG
HUDF~J033236.83-274558.0 at $z=3.797$ (its celestial coordinates are
{\it RA=03:32:36.85}, {\it DEC=-27:45:57.6}). This pair has already been discussed by
V10a, who argued that the LyC candidate is most likely a low redshift
interloper. Here we add another piece of evidence in support of this
interpretation. We have complemented the ACS optical photometry, which takes
advantage of the HUDF images (Beckwith et~al. 2006), with the new WFC3/IR data
obtained as part of the CANDELS project (Koekemoer et~al. 2011; Grogin
et~al. 2011) in the Y (F105W), J (F125W) and H (F160W) bands.  
The SED of the $Uem$ from 0.5 to 1.6 $\mu m$ is blue, essentially consistent with a flat
spectrum, i.e., $f_{\nu}\propto \nu^0$, except in the $J$ band where the source
is $\simeq$ 0.43 mags brighter (at 10-sigma), namely $U$=28.63$\pm$0.2,
$B_{435}$=28.59$\pm$0.05, $V_{606}$=28.31$\pm$0.03, $i_{775}$=28.31$\pm$0.03,
$z_{850}$=28.23$\pm$0.05, $Y$=28.31$\pm$0.07, $J$=27.88$\pm$0.04 and
$H$=28.32$\pm$0.06. Using the {\it HST} photometry alone, with or without
the U band, the photometric redshift remains unchanged, i.e., $z_{phot} \simeq 3.8$.

At the redshift of the LBG, $z= 3.797$, the $J$ band probes the
rest--frame spectral range $\lambda\lambda\sim 2295, 2918$ \AA\, where there
are no strong emission lines, except for the possible
[Mg\,{\sc ii}]~2797,2803 P--Cygni feature, which, however, has
not been observed to yield any net emission in star--forming galaxies at
redshift up to $z\sim 2$ (e.g., Weiner et~al. 2009; Rubin et~al. 2010;
Giavalisco et~al. 2011). The simplest and, in our view, most convincing
interpretation is that the U--band source is at lower redshift and the J band
flux is boosted by an emission line (or more than one), most likely
[O\,{\sc ii}] or H$\beta$ and [O\,{\sc iii}].

All photometric redshift solutions lie at $z<3$.
The required rest--frame equivalent width of a
single line to boost the $J$--band flux of 0.4 magnitude is 670\AA~if
H$\alpha$ at $z\simeq0.9$ ($L_{H\alpha}=1.5 \times 10^{40} erg/s$), or
383\AA~if [O\,{\sc ii}]3727 at $z\simeq2.5$ ($L_{[OII]}=1.9 \times 10^{41} erg/s$).

Also the complex of the three lines H$\beta$, [O\,{\sc iii}] 4959 and
[O\,{\sc iii}] 5007 can fall in the suitable wavelength interval. In the case of
H$\alpha$ we would detect [O\,{\sc ii}]3727, H$\beta$, and [O\,{\sc iii}] 4959
-- 5007 in the spectrum of the LBG, that is not the case.  Similarly, in the
case of H$\beta$ and [O\,{\sc iii}] complex, the [O\,{\sc ii}]3727 line would
be present in the spectrum of the LBG at $\sim$ 9700\AA. However in this
latter case it may be too faint to be detected.  Therefore we exclude the
H$\alpha$ possibility only, and keep the other two configurations.

Given the response curve of the WFC3/Y/J/H filters, in order to have the above
complex of three lines or the single [O\,{\sc ii}]3727 line in the $J$ filter
only (green shaded region of Figure~\ref{fig7}), the redshift of the source would be
$1.45<z<1.8$ or $2.25<z<2.7$, respectively.

As mentioned above, the required rest-frame equivalent widths are high (larger
than 300\AA), however we note that ultra-strong emission line (compact)
galaxies with these features have been recently detected (Kakazu et al. 2007;
Hu et al. 2009; Izotov et al. 2011 and more recently van der Wel et al. 2011).  
Our source would extend their findings
to a more extreme limit of luminosity, i.e.,  a rest-frame absolute $B_{435}$--band
magnitude of -16.86.  Going a bit more into the details, we note that the SED
shows a significant decrease of flux in the $U$ and $B_{435}$ bands (see
Figure~\ref{fig7}), that would $not$ be consistent with the redshift $z\sim1.6$, which
corresponds to $\lambda>1200$\AA.  Conversely, in the case of [O\,{\sc ii}]3727
at $z \sim 2.5$, these bands are affected by the IGM and the Lyman limit
attenuation. Indeed the best photometric redshift solution
is $zphot \sim 2.5$.
Therefore we argue that the compact source describe here is at $z<3$ and a
probable strong Oxygen emitter (similar to those recently identify in the CANDELS survey
by van der Wel et al. 2011), that mimic a LyC emission of the background
LBG.

\clearpage

\end{document}